\documentclass[12pt]{article}
\usepackage[margin=2cm]{geometry}  % Adjust the margin here
\usepackage[authoryear,round]{natbib}
\usepackage{url}

\usepackage{authblk}
\usepackage{physics}
\DeclareMathOperator{\Div}{Div}
\usepackage{graphicx} % Required for inserting images
\usepackage[hidelinks]{hyperref}
\usepackage{xcolor}
\usepackage{caption}
\usepackage{subcaption}
\usepackage{amsmath, amssymb, amsthm}
\usepackage{mathtools} % extension package to amsmath
\usepackage{soul}
\usepackage{siunitx} % for si units
\usepackage{enumerate}% http://ctan.org/pkg/enumerate
\usepackage{enumitem}% http://ctan.org/pkg/enumitem
\usepackage{float} % for placing figures
\usepackage{tikz} %for tikz images such as grids for flows
\usetikzlibrary{matrix}

\newcommand{\R}{\mathbb{R}}

\theoremstyle{definition}
\newtheorem{definition}{Definition}[section]

\title{{Mechanical Comparison of Arrangement Strategies for Topological Interlocking Assemblies}}
\author[1]{Tom~Goertzen$^*$}
\affil[1]{RWTH Aachen University, Chair of Algebra and Representation Theory, Pontdriesch 10-16, 52062 Aachen, Germany}

\author[2]{Domen~Macek$^*$}
\affil[2]{RWTH Aachen University, Institute of Applied Mechanics, Mies-van-der-Rohe-Str.\ 1, 52074 Aachen, Germany}

\author[1]{Lukas~Schnelle$^*$}

\author[1]{Meike~Weiß$^*$}

\author[2]{Stefanie~Reese}

\author[2]{Hagen~Holthusen}

\author[1]{Alice~C.\ Niemeyer}

\providecommand{\keywords}[1]{\textbf{\textit{Keywords:}} #1}

\date{}

\begin{document}

\maketitle
\def\thefootnote{*}\footnotetext{These authors contributed equally to this work.}
\begin{abstract}

Topological Interlocking assemblies are arrangements of blocks kinematically constrained by a fixed frame, such that all rigid body motions of each block are constrained only by its permanent contact with other blocks and the frame. In the literature several blocks are introduced that can be arranged into different interlocking assemblies. In this study we investigate the influence of arrangement on the overall structural behaviour of the resulting interlocking assemblies. 
This is performed using the Versatile Block, as it can be arranged in three different doubly periodic ways given by wallpaper symmetries. Our focus lies on the load transfer mechanisms from the assembly onto the frame. For fast a priori evaluation of the assemblies we introduce a combinatorial model called Interlocking Flows. To investigate our assemblies from a mechanical point of view we conduct several finite element studies. These reveal a strong influence of arrangement on the structural behaviour, for instance, an impact on both the point and amount of maximum deflection. The results of the finite element analysis are in very good agreement with the predictions of the Interlocking Flow model. Our source code, data and examples are available under \href{https://doi.org/10.5281/zenodo.10246034}{https://doi.org/10.5281/zenodo.10246034}.

\end{abstract}


\newpage

\section{Introduction}
The aim of resource efficiency and resource savings drives us to optimise not only the recyclability of
everyday consumer goods but also the recyclability of components in the  construction industry.
Components are typically manufactured monolithically tailored to a specific application and consist of high performance composites. 
They require separation for recycling, which in most cases consumes additional energy. This raises the question of how to achieve resource efficiency without the need for recycling.
One possible solution is to start at an earlier stage, namely to design buildings with reusable components. 
To achieve reusability, a transition from a monolithic approach to a modular design is required.
Reusable components offer the potential of being assembled into different load-bearing structures.
For example, components consisting of individual blocks that kinetically constrain each other and display structural load-bearing behaviour are particularly desirable.

The idea of building mortarless structures from blocks that kinematically constrain each other has been known for a long time (see Section \ref{sec:lit-rev}). We are particularly interested in topological interlocking assemblies which give rise to planar mortarless structures. The structural behaviour of an interlocking assembly of tetrahedra has first been investigated in \cite{dyskin_new_2001}.

Mathematically, a topological interlocking assembly can be defined as an arrangement of blocks that are in contact with each other together with a frame such that, if the frame is fixed, any non-empty finite subset of blocks of the arrangement is prevented from moving. This restriction of movement is enforced by the neighbours of any block and thus captures the aim of having blocks that kinetically constrain each other. While in the well known tetrahedral interlocking assembly the individual tetrahedra can only be assembled in one way, other blocks exist that can be arranged into topological interlocking assemblies in a number of ways (see \citet{dyskin_fracture_2003} and \citet[Lemma 1]{bridges23}).

In this contribution, we investigate the influence of the arrangement on the structural behaviour of a topological interlocking assembly. Our investigations focus on a specific block, called the Versatile Block (see Section \ref{sec:vers-block}), that allows many different arrangements. Here we focus on three different symmetric assemblies that can be used as plate-like structures, called planar assemblies.
The results of this paper show that these arrangements of the Versatile Block into planar topological interlocking assemblies display surprisingly different mechanical performance.

As bending loads are most relevant for plates, these are perfectly suited to understand the load transfer
mechanisms of topological interlocking assemblies. To evaluate the three considered interlockings in terms of load-bearing behaviour and serviceability, we compare the equivalent stresses and deflections of the topological interlocking assemblies with the results of a monolithic plate (Section \ref{sec:mech-inv}).

Finally, we introduce a combinatorial tool (Section \ref{sec:comb-tool}), which we call `Interlocking Flows', that allows a fast prediction of how load transfer onto a frame occurs. The pre-evaluation obtained by the combinatorial tool of the assemblies we consider is consistent with the results of the FEM analyses we conducted (Section \ref{sec:mech-inv}).

\section{Literature Review}\label{sec:lit-rev}

\subsection{Design Principles and Applications}
The concept of topological interlocking assemblies (TIA), also known as topological interlocking materials (TIM) or topological interlocking structures (TIS), has a long history. It is related to the concept of masonry and the idea of building flat vaults. Early patents and concepts of blocks that admit a topological interlocking assembly can be found in the work of Abeille and of Truchet in \cite{abeille_memoire_1735} as well as Frézier, who generalizes the work of Abeille and Truchet, \cite{frezier_theorie_1738}.
The block proposed by Abeille can be viewed as a truncated tetrahedron. Glickman proposes a paving block related to an assembly of tetrahedra \cite{glickman_g-block_1984}. \cite{dyskin_new_2001} initiate an investigation of topological interlocking assemblies as a novel material design concept and coin the term `topological interlocking' in \cite{dyskin_toughening_2001}. Moreover, they show that all platonic solids give rise to topological interlocking assemblies \cite{dyskin_topological_2003} and describe a method for constructing TIA with convex blocks \cite{kanel-belov_interlocking_2010}. 
Osteomorphic type blocks, which are introduced in \cite{dyskin_fracture_2003}, can also be assembled in various non-planar ways and this versatility gives rise to applications in civil engineering, see \cite{dyskin_fracture_2003,yong_utilisation_2011,javan2016}. Other methods for generating TIA linked to Voronoi tesselations are proposed in \cite{subramanian_delaunay_2019,akleman_generalized_2020,mullins_voronoi_2022}. Voronoi tesselations are naturally linked to crystallographic groups as certain (convex) Voronoi cells yield fundamental domains for such groups and thus space-filling structures, for example space-filling `VoroNoodles' \cite{ebert_voronoodles_2023}.

A general method for constructing planar TIA based on non-convex fundamental domains of a crystallographic group is introduced in \cite{topological22} and for non-planar TIA in \cite{spherical_interlocking}. One block, called the \emph{Versatile Block}, generated by the method given in \cite{topological22} has an additional property of versatility similar to the osteomorphic block:  the Versatile Block can be arranged into TIA in infinitely many ways, three of which are invariant under the symmetries of a planar crystallographic group \cite{bridges23}.
Recent overviews of design principles and applications related to TIA are given in \cite{dyskin_topological_2019,estrin_design_2021}.

\subsection{Mechanical and Experimental Investigations}

In this paper, the focus lies on the mechanics of certain TIA based on the Versatile Block \cite{bridges23,topological22}. For this endeavour, the simulation software ABAQUS is applied. Different investigations in the literature focus on TIA made of several materials such as brittle materials (ceramics, concrete), metals (aluminium, steel) and different kinds of plastic. TIA based on ceramics are, for instance, investigated in \cite{krause_mechanical_2012}, where the authors compare fracture toughness between monolithic ceramic plates and assemblies of osteomorphic ceramics parts. In \cite{mirkhalaf_simultaneous_2018}  a parameter study comparing several convex blocks based on regular square and hexagon tessellations is conducted. Several mechanical experiments and simulations studying the behaviour of TIA with osteomorphic type concrete blocks are conducted in \cite{javan2017_experiments,javan_impact_2018,Javan_2020_rubber}. Isotropic linear–elastic materials such as aluminium and steel are investigated in  \cite{dugue_indentation_2013} using finite element and discrete element methods. In \cite{schaare_point_2008,schaare_damping_2009} TIA based on aluminium and steel cubes are experimentally and numerically studied. Mechanical properties of parameterised TIA based on `VoroNoodles' are analysed in \cite{ebert_voronoodles_2023}. Mechanical properties of TIA made with plastics such as Acrylonitrile butadiene styrene (ABS) are investigated in several works. In \cite{weizmann_effect_2021} assemblies of convex ABS blocks are generated and compared based on a parameter study with angles and different tessellations. Impact mechanics of TIA with tetrahedra are studied in \cite{feng_impact_2015}. Various parametric studies based on scaling and convex blocks coming from various tessellations are conducted in \cite{short_scaling_2019,kim_mechanics_2021,williams_mechanics_2021}.
Investigations on effects of Young’s modulus and the friction coefficient on the structural mechanics of TIA are investigated in \cite{koureas_failure_2022,feldfogel_scaling_2023,feldfogel_failure_2024}. In \cite{koureas_beam-like_2023} the effect of non-planar block geometry in the context of beam-like structures is studied. In \cite{ullmann_deflection_2023} a comparative study on the deflection limit of slab-like assemblies  monolithic slabs is conducted.

\section{Symmetric topological interlocking assemblies}
\subsection{Topological Interlocking}
A topological interlocking assembly (TIA) can be defined as an arrangement of blocks that are in contact with each other together with a frame such that, if the frame is fixed, any non-empty finite subset of blocks of the arrangement is prevented from moving.
% Many topological interlocking assemblies have been proposed in the literature \cite{some-examples} and several display desirable properties in terms of resistance to crack propagation, sound absorption, or load distribution \cite{TODO,Estrin}. 
We focus on planar topological interlocking assemblies, in which blocks are arranged between two parallel planes in 3D-space. In this scenario the frame can consist of the blocks on the perimeter.
In this paper we consider topological interlocking assemblies constructed from copies of the same block and investigate the question what effect the arrangement of these blocks has on the load bearing behaviour of the entire assembly.
Even with the restriction of only using copies of the same block a large number of planar topological interlocking assemblies can exist (possibly exponentially many, see \cite[Lemma 1]{bridges23}).
In this paper, we consider the even more restrictive case of interlocking assemblies displaying a wallpaper symmetry, more in Section \ref{subs:ass-vers-block}.

\subsection{The Versatile Block}\label{sec:vers-block}
\begin{figure}[H]
    \centering
    \begin{subfigure}[c]{0.3\textwidth}
        \centering
        \includegraphics[width=\textwidth]{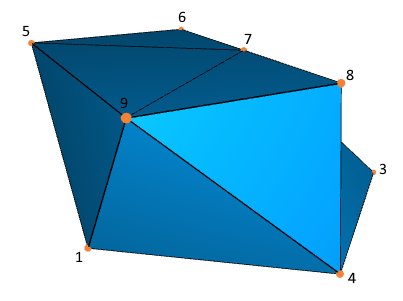}
    \end{subfigure}
    \hfill
    \begin{subfigure}[c]{0.3\textwidth}
        \centering
        \includegraphics[width=\textwidth]{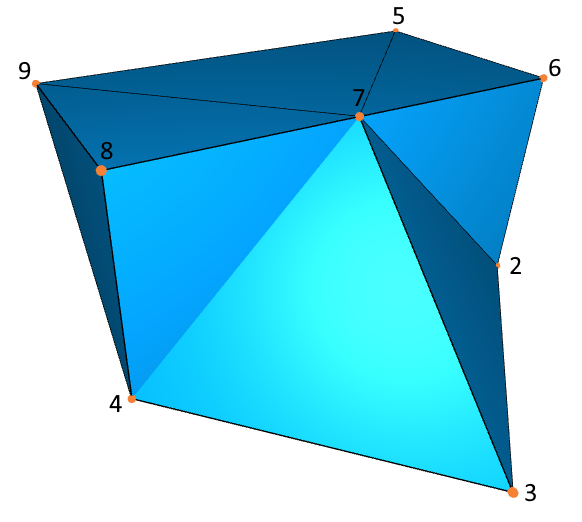}
    \end{subfigure}
    \hfill
    \begin{subfigure}[c]{0.3\textwidth}
        \centering
        \includegraphics[width=\textwidth]{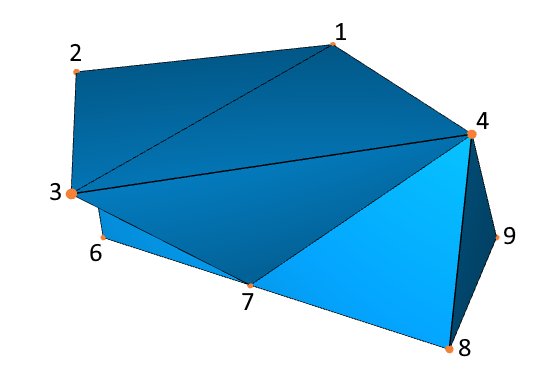}
    \end{subfigure}
    \label{fig:versatile-block-single}
    \caption{Three views of the Versatile Block}
\end{figure}

The Versatile Block, called $B$, is a polyhedron embedded in $\R^3$. It consists of vertices, called $B_0$, edges, called $B_1$, and triangular faces, called $B_2$, and was first defined in \cite{topological22}. 
More precisely, $B$ has $9$ vertices, $21$ edges and $14$ faces. Let $B_0=\{v_1,\ldots,v_9 \, \}$ denote the vertices of $B$. Then the edges $B_1$ are the following 2-subsets of vertices:
\begin{align*}
    B_1 := \{ \, &\{ v_1, v_2 \}, \, \{ v_1, v_3 \}, \, \{ v_1, v_4 \},\, \{ v_1, v_5 \},\, \{ v_1, v_9 \}, \, \{ v_2, v_3 \},\, \{ v_2, v_5 \}, \\
    & \{ v_2, v_6 \}, \, \{ v_2, v_7 \}, \, \{ v_3, v_4 \}, \, \{ v_3, v_7 \}, \, \{ v_4, v_7 \},\, \{ v_4, v_8 \},\, \{ v_4, v_9 \},\\
    & \{ v_5, v_6 \},\, \{ v_5, v_7 \}, \, \{ v_5, v_9 \}, \,\{ v_6, v_7 \}, \, \{ v_7, v_8 \}, \, \{ v_7, v_9 \}, \, \{ v_8, v_9 \} \, \}
\end{align*} 
and the triangular faces $B_2$ are uniquely described by the following $3$-subsets of vertices:
\begin{align*}
 B_2 := \{\,& \{v_1, v_2, v_3\},\, \{v_1, v_2, v_5\}, \, \{v_1, v_3, v_4\}, \,\{v_1, v_4, v_9\},\, \{v_1, v_5, v_9\}, \,\{v_2, v_3, v_7\}, \, \{v_2, v_5, v_6\},\\
&\{v_2, v_6, v_7\}, \, \{v_3, v_4, v_7\}, \, \{v_4, v_7, v_8\}, \, \{v_4, v_8, v_9\}, \, \{v_5, v_6, v_7\}, \, \{v_5, v_7, v_9\}, \,\{v_7, v_8, v_9\}\,\}.
\end{align*}
The embedding of the Versatile Block into $\R^3$ has the following coordinates, see \cite{bridges23}:
\begin{alignat*}{3}
&v_1 = (0, 0, 0) ,\, v_2 = (1, 1, 0) , \, v_3 = (2, 0, 0) ,\, v_4 = (1, -1, 0) , \\
&v_5 = (0, 1, 1) , \, v_6 = (1, 1, 1) , \, v_7 = (1, 0, 1) ,\, v_8 = (1, -1, 1) ,\, v_9 = (0, -1, 1).
\end{alignat*} 
The pair $\{\,\{v_1, v_2, v_3\}, \, \{v_1, v_3, v_4\}\, \}$ of faces forms a square in the plane $z=0$ with area 2, whereas the set $\{\, \{v_5, v_6, v_7\},\, \{v_5, v_7, v_9\}, \, \{v_7, v_8, v_9\} \,\}$ of faces forms a rectangle in the plane $z=1$ of the same area. In fact, the intersection of $B$ with any plane $z=a$ with $a\in[0,1]$ yields a polygon with area $2$.
We call the plane $z=0$ the \emph{bottom plane} and the plane $z=1$ the \emph{top plane}. Note that all vertices of the Versatile Block lie in either the top or bottom plane.\\

The Versatile Block can be used to create many different planar assemblies sandwiched between the bottom and top plane, such that each block is placed in such a way, that the square is in the bottom plane and the rectangle is in the top plane, see \cite[Lemma 1]{bridges23}.
In this paper we are interested in planar topological interlocking assemblies of copies of the Versatile Block, that remain unchanged under certain rotations, reflections and translations. For this we need to introduce a formal way to describe these symmetries.

\subsection{Wallpaper groups}\label{subs:wallpaper-grp}
Wallpaper groups can be seen as a mathematical formulation of symmetries of certain doubly periodic repeating patterns in a $2$-dimensional plane. They are also known as $2$-dimensional crystallographic groups and contain rotations, reflections and translations respecting the repeated pattern.
For the repetition of the pattern we need a translation which describes the offset of the pattern, denoted as a vector in $\R^2$. Furthermore, a rotation or reflection can be described by a matrix $$\begin{pmatrix}cos(\theta) &-sin(\theta) \\ sin(\theta) &cos(\theta)\end{pmatrix}\text{ or }  \begin{pmatrix}cos(\theta) &sin(\theta) \\ sin(\theta) &-cos(\theta)\end{pmatrix}$$ respectively, where $\theta$ is the angle. A reflection means, we mirror at a plane through the origin with a certain angle. It turns out, that $\theta$ can only have the values $0^\circ, 60^\circ, 90^\circ, 120^\circ,$ and $180^\circ$ (see crystallographic restriction theorem \cite[Chapter 25]{armstrong1997groups}). A single symmetry of a repeating pattern can therefore be described as pair $(M, v)$ where $M$ is a rotation or reflection matrix and $v$ is an (offset) vector as before. We call such an object an \emph{isometry}. 
An isometry $(M, v)$ acts on $\R^2$ as the function 
$$f_{(M,v)} : \R^2 \to \R^2, x = \begin{pmatrix}x_1\\ x_2\end{pmatrix} \mapsto M \cdot \begin{pmatrix}x_1\\ x_2\end{pmatrix} + v,$$
whereas the product of two isometries $(M_1,v_1), (M_2,v_2)$ is defined as follows: 
$$
(M_2, v_2) \circ (M_1, v_1) \coloneqq (M_2 \cdot M_1, M_2 \cdot v_1 + v_2)
$$
with $\cdot$ the matrix-vector multiplication. Every wallpaper group is generated by a finite set of isometries, see \cite[Def 2.1]{szczepanski2012geometry}. That means elements of the group are products of these isometries. To be more precise we define a wallpaper group along the same lines as \cite[Chapter 25]{armstrong1997groups} as follows:

\begin{definition} \ \\[0.05cm]
    Let $\mathrm{E}(2)$ denote the group of isometries of the Euclidean plane $\R^2$. A subgroup 
    $$\Gamma \coloneqq \langle \, (M_1, v_1) , \dots , (M_r, v_r)\, \rangle \subseteq \mathrm{E}(2)$$
    such that $M_i$ is a rotation or reflection matrix and $v_i \in \R^2$ for $1\leq i \leq r$ is called \emph{wallpaper group} if
    \begin{enumerate}[label=(\roman*)]
        \item the set $\{ v_1, \dots , v_r\}$ contains two linearly independents vectors and
        \item there are only finitely many matrices that can be written as $M = M_{i_1} \cdot \dots \cdot M_{i_k}$ with $M_{i_j} \in \{ M_1, \dots, M_r\}$, $1 \leq j \leq k$ (i.e. the $M_i$ span a finite \emph{point group}).
    \end{enumerate}
\end{definition}

In this definition, $(i)$ ensures that we can find a smallest area $D \subseteq \R^2$ (called fundamental domain) such that for any point $x$ in the plane there is a point $y\in P$ and an isometry in $\Gamma$ that maps $x$ to $y$.
$(ii)$ ensures that we obtain a repeating pattern, which was our goal. It turns out, that with this definition it can be shown that there are only $17$ wallpaper groups (up to isomorphism), see \cite[Chapter 26]{armstrong1997groups}. 

In this paper we only consider the following three wallpaper groups:

$$p1 \coloneqq \left< 
\left( \begin{pmatrix} 1 &0 \\ 0 &1\end{pmatrix}, \begin{pmatrix} 1 \\ -1 \end{pmatrix}\right), 
\left( \begin{pmatrix} 1 &0 \\ 0 &1\end{pmatrix}, \begin{pmatrix} 1 \\ 1 \end{pmatrix}\right) \right>.$$        
        The group $p1$ can be characterised by only allowing translations as both matrices are the identity.
$$pg\coloneqq \left< 
\left( \begin{pmatrix} 0 &-1 \\ -1 &0\end{pmatrix}, \begin{pmatrix} 2 \\ 0 \end{pmatrix}\right), 
\left( \begin{pmatrix} 1 &0 \\ 0 &1\end{pmatrix}, \begin{pmatrix} 1 \\ 1 \end{pmatrix}\right) \right>.$$
        In the group $pg$ we gain a (glide-)reflection in the first generator. After applying this matrix twice it becomes the identity and therefore the fundamental domain $D$ can be oriented in two ways.
$$p4\coloneqq \left< 
\left( \begin{pmatrix} 0 &1 \\ -1 &0\end{pmatrix}, \begin{pmatrix} 0 \\ 2 \end{pmatrix}\right), 
\left( \begin{pmatrix} 0 &-1 \\ 1 &0\end{pmatrix}, \begin{pmatrix} 0 \\ -2 \end{pmatrix}\right), \left( \begin{pmatrix} -1 &0 \\ 0 &-1\end{pmatrix}, \begin{pmatrix} 0 \\ 0 \end{pmatrix}\right) \right>$$
        Compared to the one reflection in $pg$ we have a rotation associated to both generators in the group $p4$. However, due to the rotation matrices becoming the identity after being applied four times the fundamental domain $D$ can be oriented in exactly four ways.

Here, we follow the international crystallographic notation \cite{itc2016} for the names of the wallpaper groups.%, where $p$ stands for primitive, $g$ for glide-reflection and the number after $p$ corresponds to a rotation, see TODO international tables.

\subsection{Planar Assemblies of the Versatile Block}\label{subs:ass-vers-block}
\begin{figure}
    \centering
    \begin{subfigure}[b]{0.32\textwidth}
        \centering
        \includegraphics[width=\textwidth]{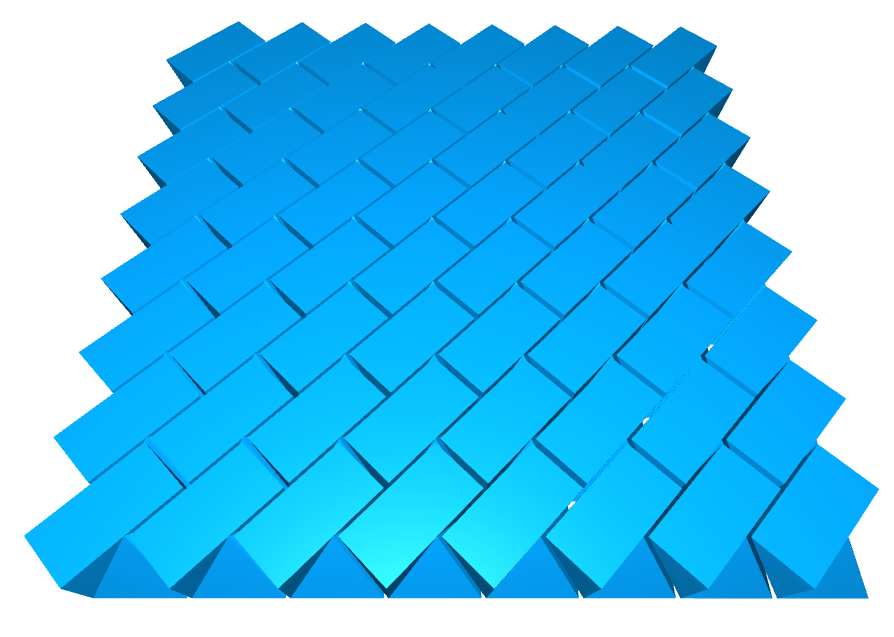}
        \caption{$p1$}
        \label{fig:p1}
    \end{subfigure}
    \hfill
    \begin{subfigure}[b]{0.32\textwidth}
        \centering
        \includegraphics[width=\textwidth]{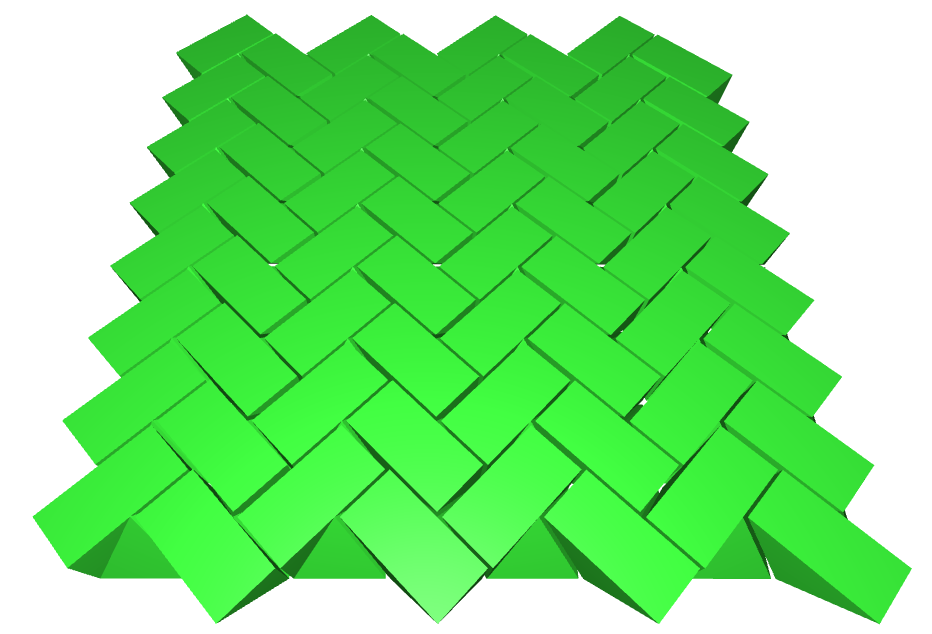}
        \caption{$pg$}
        \label{fig:pg}
    \end{subfigure}
    \begin{subfigure}[b]{0.32\textwidth}
        \centering
        \includegraphics[width=\textwidth]{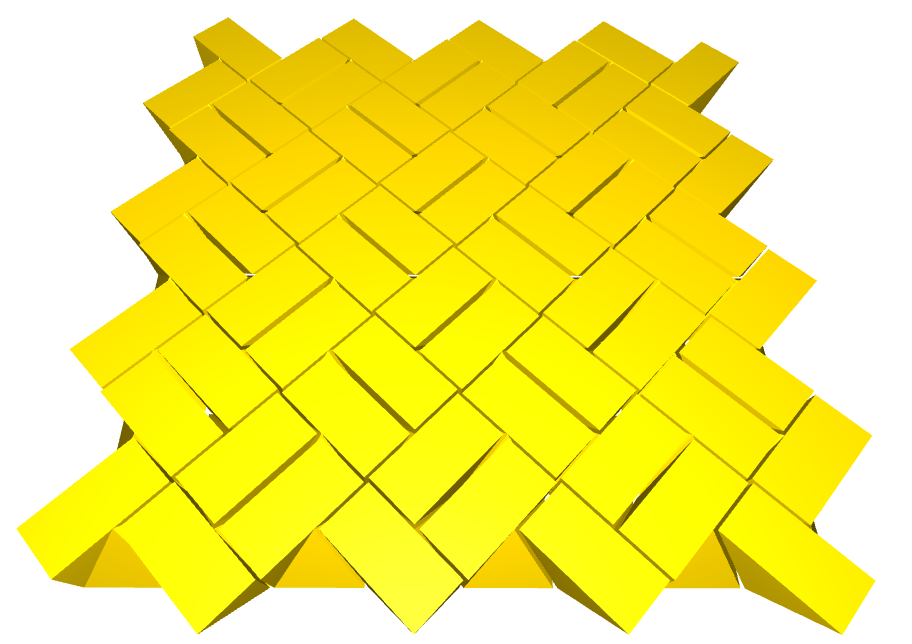}
        \caption{$p4$}
        \label{fig:p4}
    \end{subfigure}
    \caption{Planar interlocking assemblies of size $9 \times 9$ with a gap of $0.1$ between blocks, generated by a their respective wallpaper group}
    \label{fig:versatile-ass}
\end{figure}

\begin{figure}
    \centering
    \begin{subfigure}[b]{0.32\textwidth}
        \centering
        \includegraphics[width=\textwidth]{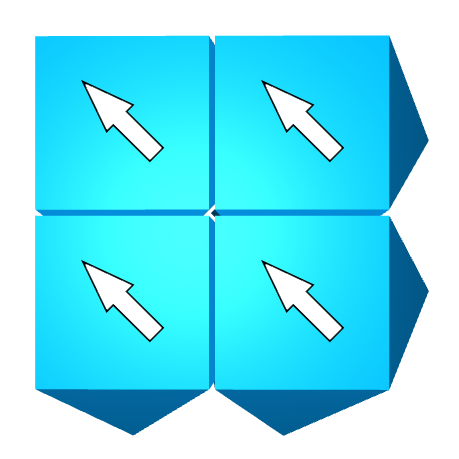}
        \caption{$p1$}
        \label{fig:p1}
    \end{subfigure}
    \hfill
    \begin{subfigure}[b]{0.32\textwidth}
        \centering
        \includegraphics[width=\textwidth]{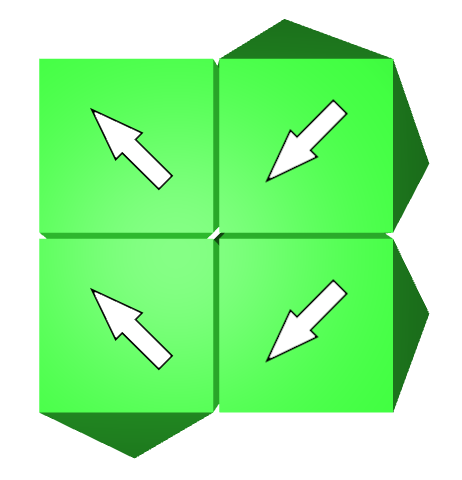}
        \caption{$pg$}
        \label{fig:pg}
    \end{subfigure}
    \begin{subfigure}[b]{0.32\textwidth}
        \centering
        \includegraphics[width=\textwidth]{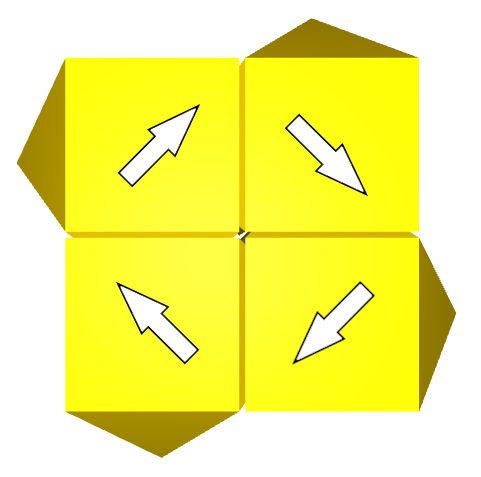}
        \caption{$p4$}
        \label{fig:p4}
    \end{subfigure}
    \caption{View of the bottom-plane with arrows towards vertex $3$ to indicate the orientation}
    \label{fig:ass-bott}
\end{figure}
\begin{figure}
    \centering
    \begin{subfigure}[b]{0.32\textwidth}
        \centering
        \includegraphics[width=\textwidth]{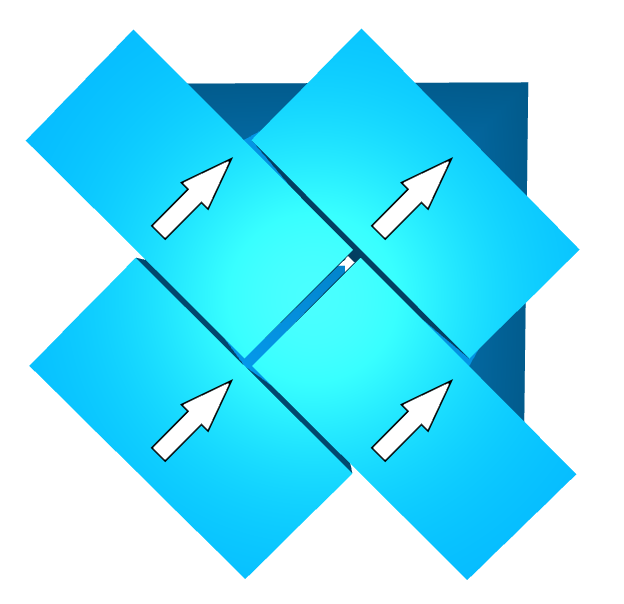}
        \caption{$p1$}
        \label{fig:p1}
    \end{subfigure}
    \hfill
    \begin{subfigure}[b]{0.32\textwidth}
        \centering
        \includegraphics[width=\textwidth]{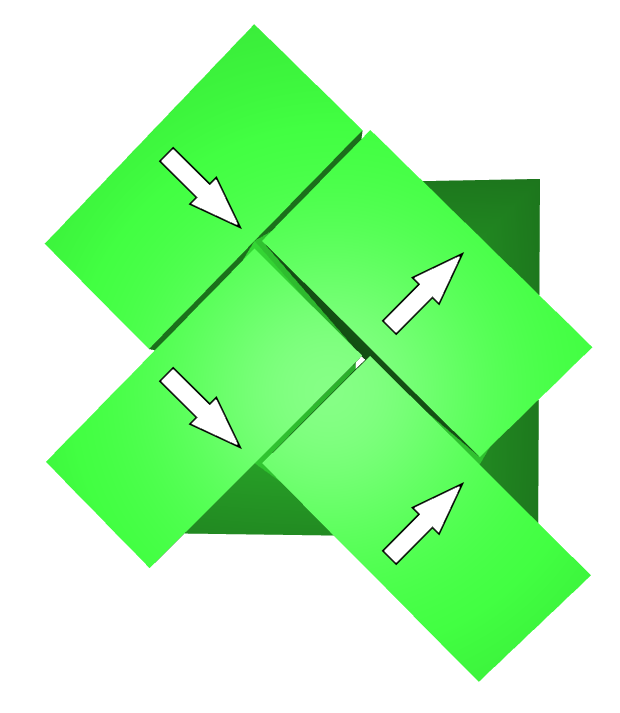}
        \caption{$pg$}
        \label{fig:pg}
    \end{subfigure}
    \begin{subfigure}[b]{0.32\textwidth}
        \centering
        \includegraphics[width=\textwidth]{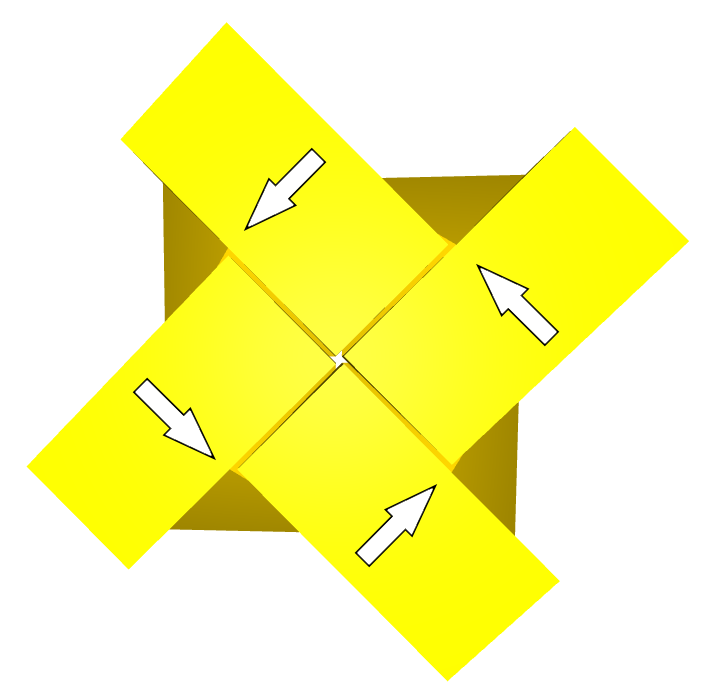}
        \caption{$p4$}
        \label{fig:p4}
    \end{subfigure}
    \caption{View of the top-plane with arrows towards vertex $3$ to indicate the orientation}
    \label{fig:ass-top}
\end{figure}
The Versatile Block is constructed in a way such that we obtain infinite planar topological interlocking assemblies with wallpaper symmetries $p1$, $pg$ or $p4$ if the blocks are assembled in the following way:\\
In the bottom plane we consider the region $P$ as the square of the Versatile Block. We can apply an isometry on the square in the bottom plane, and keep the coordinates of the rectangle relative to the ones of the square,  i.e. the isometry acts on the first two coordinates the usual way and does not change the last coordinate. Formally, this means that we extend an isometry of the Euclidean plane $\mathbb{R}^2$ to Euclidean three-space $\mathbb{R}^3$ for a given tuples $(M,v)$ to a map $$ \hat{f}_{(M,v)} : \R^3 \to \R^3, x = \begin{pmatrix}x_{1}\\
x_{2}\\
x_{3}
\end{pmatrix}\mapsto\begin{pmatrix}M\cdot\begin{pmatrix}x_{1}\\
x_{2}
\end{pmatrix}+v\\
x_{3}
\end{pmatrix}.$$ Notice, that an isometry always keeps distances of any two vectors the same. Thus, after applying a non-trivial isometry the result is a Versatile Block rotated and translated as a whole without changing its shape and size,  more in \cite{topological22} and \cite{bridges23}.
The three planar interlocking assemblies generated by the wallpaper groups $p1$, $pg$, and $p4$ are shown in Figure \ref{fig:versatile-ass}.

\section{Mechanical investigation of interlocking assemblies}\label{sec:mech-inv}

Our primary goal is to utilise numerical simulations to predict and characterise the mechanical behaviour of topological interlocking assemblies. By doing so, we aim to gain valuable insights into the intricate mechanisms of topological interlocking assemblies. Specifically, our focus lies in understanding the dynamic interaction between the blocks within a planar assembly when subjected to transverse loading. We examine how external forces are transferred to the frame that holds the assembly together. To further improve our understanding of the topological interlocking assembly from a mechanical point of view, we conduct a comparative analysis of the bear-loading behaviour between the assemblies and monolithic plates of the same geometry, specifically evaluating the maximum deflection and distribution of stresses. \\

\subsection{Problem formulation}
Mechanically a topological interlocking assembly consists of a set of independent bodies, which are arranged in the space and held together by the fixed, undeformable frame. In our case, the bodies are the Versatile Block.
The bodies are not tied together and are considered to be deformable solids. Only the external forces (no displacements) are prescribed to the bodies in the assembly. The Versatile Blocks interact with each other only by the contact, which is unknown a priori and can change over time. The analysis does not consider friction between the blocks, allowing for a pure interlocking effect. Both the contact forces and displacements on the contact boundary of each block are not prescribed. Thus, the geometry of the blocks “constrains” the relative motion between them. Contact interaction can be mathematically interpreted as a set of nonlinear boundary conditions (see \cite{laursen2003computational} and \cite{wriggers2006computational}).

The reference configuration $\Omega_0^{(k)} \subset \mathbb{R}^3$ of a body $k$ denotes the domain occupied by all material points $\vb{X}^{(k)}$ at time $t=0$. The changed positions $\vb{x}^{(k)}$ of a material point at a certain time $t$ are described by the current configuration $\Omega_t^{(k)} \subset \mathbb{R}^3$. The displacement of a material point is described by $\vb{u}^{(k)}(\vb{X}^{(k)},t) = \vb{x}^{(k)}(\vb{X}^{(k)},t) - \vb{X}^{(k)}$.
The boundary of each body $\partial \Omega_0^{(k)}$ is decomposed into three sets: $\Gamma_{\sigma}^{(k)}$ representing the Neumann boundary (tractions $\bar{\vb{t}}_0^{(k)}$ are given), $\Gamma_{u}^{(k)}$ representing the Dirichlet boundary (displacements $\bar{\vb{u}}^{(k)}$ are given), and $\Gamma_{c}^{(k)}$ representing the contact surface.
The initial boundary value problem (strong formulation) of finite deformation elastodynamics needs to be satisfied on each body:
\begin{align}\label{eq:inital-bound}
    \Div \vb{P}^{(k)} + \tilde{\vb{b}}_0^{(k)} = \rho_0^{(k)} \ddot{\vb{u}}^{(k)} \qquad &\text{in } \Omega_0^{(k)} \times [0,t],\\
    \vb{u}^{(k)} = \tilde{\vb{u}}^{(k)} \qquad &\text{on } \Gamma_u^{(k)} \times [0,t],\\
    \vb{P}^{(k)} \vb{N}^{(k)} = \tilde{\vb{t}}_0^{(k)} \qquad &\text{on } \Gamma_{\sigma}^{(k)} \times [0,t],\\
    \vb{u}^{(k)} (\vb{X}^{(k)}, 0) = \tilde{\vb{u}}^{(k)} (\vb{X}^{(k)}) \qquad &\text{in } \Omega_0^{(k)},\\
    \dot{\vb{u}}^{(k)} (\vb{X}^{(k)}, 0) = \dot{\tilde{\vb{u}}}^{(k)} (\vb{X}^{(k)}) \qquad &\text{in } \Omega_0^{(k)},\\
    g_n^{(k)}(\vb{X}^{(k)},t) \geq 0, \quad p_n^{(k)}(\vb{X}^{(k)},t) \leq 0, \quad p_n^{(k)}(\vb{X}^{(k)},t) g_n^{(k)}(\vb{X}^{(k)},t) = 0 \qquad &\text{on } \Gamma_c^{(k)} \times [0,t],\label{eq:inital-bound-con}
\end{align}
where $\vb{P}$ is the first Piola-Kirchhoff stress tensor, $\Div ()$ denotes the Lagrangian divergence, and $\vb{N}$ is the normalised unit surface normal. The contact constraints in normal direction (\ref{eq:inital-bound-con}) for frictionless contact must hold on the contact boundary $\Gamma_{c}^{(k)}$ at each time $t$. Here, $g_n$ is the gap function, and $p_n$ is the contact pressure. The true internal stress state within a body is represented by Cauchy stress tensor $\boldsymbol{\sigma}$, which has the following relation to the first Piola-Kirchhoff stress tensor $\vb{P}$ $$\boldsymbol{\sigma} = 1/J \,\vb{P} \vb{F}^T .$$ Here, $\vb{F}$ is the deformation gradient ($\vb{F}=\partial \vb{x}/\partial \vb{X}$) and $J$ is its determinant.

Contact problems can be tackled using various numerical methods, such as finite element methods (FEM), discrete element methods (DEM), and multi-body systems. The selection of the appropriate method depends on the specific nature of the problem at hand. In our analysis, we choose the finite element method, as it is well-suited for examining the deformation and stress fields arising from quasi-static problems in assemblies composed of arbitrarily shaped solids. For a more in-depth explanation of FEM, we kindly refer interested readers to, e.g., \cite{zienkiewicz2005finite}, \cite{wriggers2008nonlinear} for general theory on FEM, and \cite{wriggers2006computational}, \cite{laursen2003computational} for computational contact mechanics.

\subsection{Simulation setup}
The mechanical analyses were conducted by using the commercial finite-element software Abaqus/CAE 2022.HF1, 

in which the explicit dynamics environment was employed to obtain the quasi-static solution.

To give the analysis a more realistic face, we scaled the coordinates of the Versatile Block and the resulting assemblies, described in the previous section, by applying a diagonal matrix of the form $$\begin{pmatrix}a & 0 & 0\\0 & b & 0\\ 0 & 0 & c \end{pmatrix}$$ with $a,b,c>0$. For the experiments presented in this sections we interpret the length units as metres \SI{}{\metre} and apply the scaling matrix with $a=b=c=0.2$ and thus obtain a block of height \SI{0.2}{\metre}. In this case the side-length of the square equals $0.2\cdot \sqrt{2} \approx 0.283$ \SI{}{\metre}. Especially, a $10\times 10$ assembly of scaled Versatile Blocks in a square grid will be of size $2.83 \times 2.83$ \SI{}{\square\metre}.

Geometries were generated with in house developed code and imported into software Abaqus as .stl files. To compare the mechanical properties of these planar interlocking assemblies with some reference, a solid plate of the same dimensions was modelled as well. We considered the bodies as isotropic and linear elastic material, which properties are listed in table \ref{table:parameters}. Soft frictionless contact between all bodies was assumed and defined by exponential pressure-overclosure relationship. The displacement boundary conditions were applied by fully fixing (in all their nodes) the peripheral blocks (the bounding frame) in space. All blocks were meshed individually with 4-node tetrahedral elements. A pressure $p_0 =$ \SI{1.5}{\bar} transversely to the assembly plane was applied onto the top plane using the integrated function “smooth step”. Quasi-static loading conditions were considered. Both a damping term related to the volumetric strain rate and the square of the volumetric strain rate were considered. Material damping has been used to damp lower (mass-dependent) and higher (stiffness-dependent) frequency responses (see Table \ref{table:parameters}). To perform the quasi-static analysis efficiently, mass scaling by factor 10 was employed to increase the integration time step. All Abaqus input files can be found in \cite{zenodo_files}.

\renewcommand{\arraystretch}{1.5}
\begin{table}[h!]
  \centering
  \caption{Simulation parameters.}
  \begin{tabular}[h!]{c| c | c}
    \hline\hline
    \textbf{Parameter} &  \textbf{Value} & \textbf{Description} \\[5pt]
    \hline
    $\rho$ [\SI{}{\kilogram\per\cubic\metre}] & 7850 & Density
    \\
    \hline
    $E$ [\SI{}{\giga\pascal}] & $210$ & Young's modulus
    \\
    \hline
    $\nu$ [\SI{}{-}]& $0.3$ & Poisson's ratio
    \\
    \hline
    $\alpha$ [\SI{}{-}] & $2.0$ & Mass proportional damping
    \\
    \hline
    $\beta$ [\SI{}{-}] & $1.0 \cdot 10^{-8}$ & Stiffness proportional damping
    \\    
    \hline \hline
  \end{tabular}
  \label{table:parameters}
\end{table}

\subsection{Numerical results}

\begin{figure}[H]
    \centering
    \includegraphics[width=1.0\textwidth]{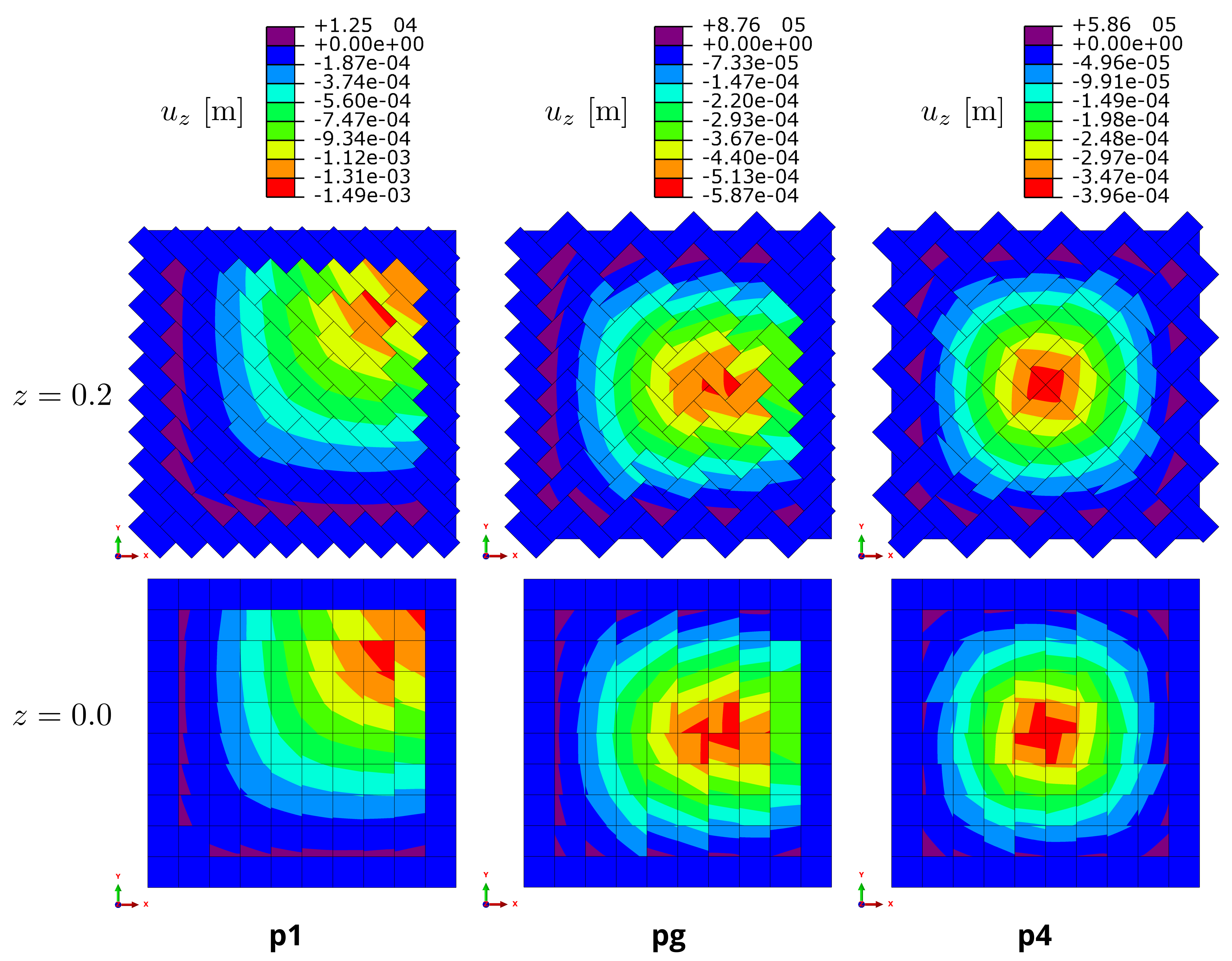}
    \caption{Displacement fields in $z$-direction of the TIA.}
    \label{fig:u3}
\end{figure}

\begin{figure}[H]
    \centering
    \includegraphics[width=1.0\textwidth]{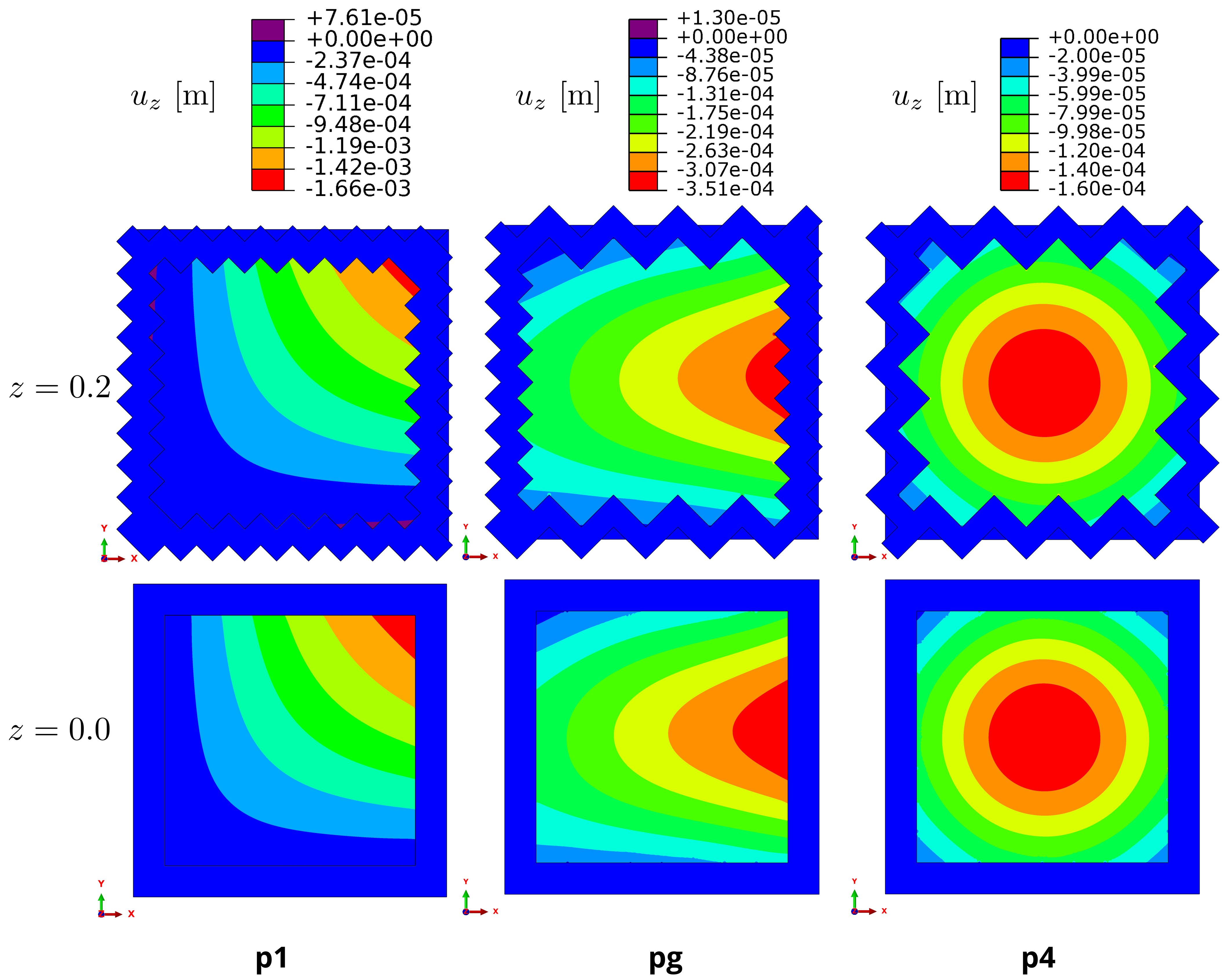}
    \caption{Displacement fields in $z$-direction of the monolithic plates.}
    \label{fig:u3_ref}
\end{figure}

\paragraph{Deformed state and maximum deflection}
Figure \ref{fig:u3} shows the displacement fields $u_z$ in $z$-direction of the three TIA. The displacement $u_z$ is shown because the deformation in $z$-direction is most dominant due to the loading direction. The deformation pattern of the interlocking assemblies follows the pattern and the direction of the monolithic plates (Figure \ref{fig:u3}) but behaves slightly differently due to the modular nature of the assemblies (Figure \ref{fig:u3_ref}). In $p1$, the displacement field spreads along the diagonal form the bottom left corner to the top right corner. The reason for this deformation that all the blocks have the same orientation. In $pg$, the deformation spreads from the left boundary in the positive $x$-direction and is limited by the top and bottom boundaries. In $p4$, the deformation is equally distributed in all directions and therefore the maximum deflection occurs in the middle. The displacements of the blocks in $p1$ at the left and bottom boundary (in purple in Figure \ref{fig:u3}) are positive, which suggests that their neighbouring blocks leverage them out. This behaviour differs from that of the solid plate. Similar differences between the solid plate and the assemblies $pg$ and $p4$ can be observed. The comparison of the maximum deflections of TIA with the solid plates shows that the $p1$ performs surprisingly better, while $pg$ and $p4$ perform about 50\% worse. The maximum deflection occurs in front of the right corner block in $p1$; slightly to the right of the centre for $pg$; and in the centre for $p4$. The results show that there is a difference in the displacement distribution between the top and bottom plane of the TIA. Overall it can be concluded that the choice of the assembly, and thus, the shape of the frame, controls the position of the maximum deflection.

\begin{figure}[H]
    \centering
    \includegraphics[width=1.0\textwidth]{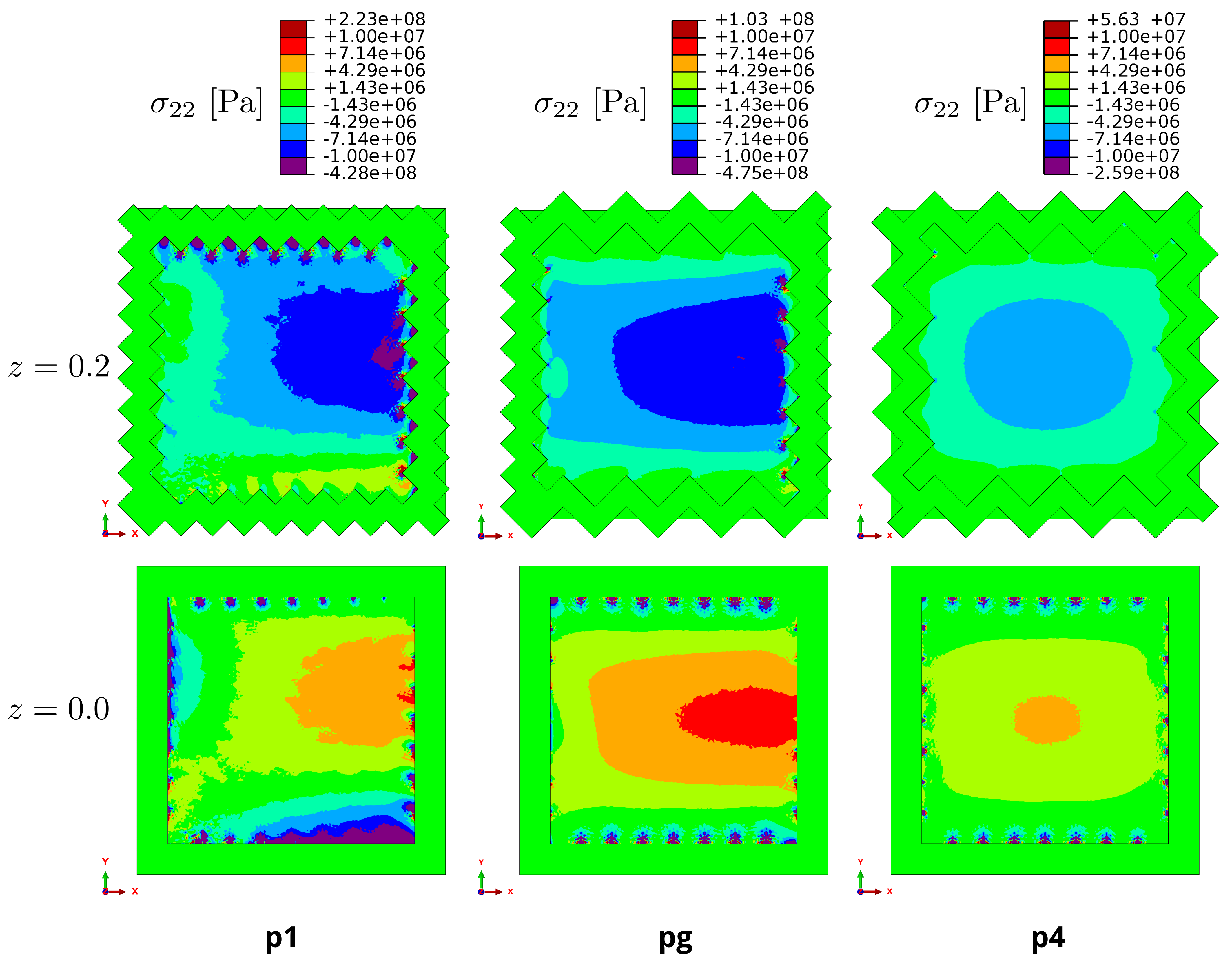}
    \caption{Normal stresses in $y$-direction of the monolithic plates.}
    \label{fig:S_22_ref}
\end{figure}

\begin{figure}[H]
    \centering
    \includegraphics[width=1.0\textwidth]{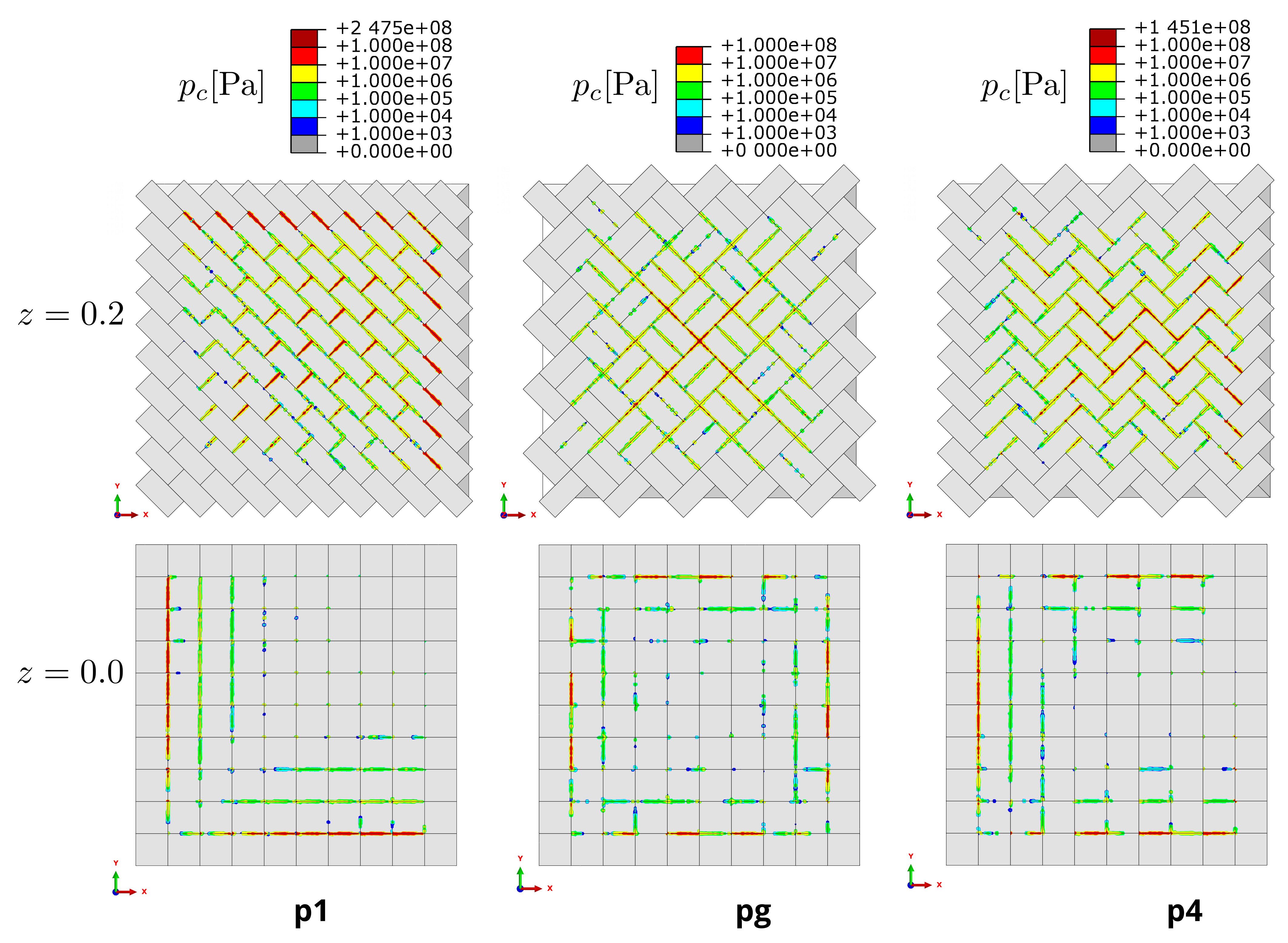}
    \caption{Contact pressure between the blocks in TIA.}
    \label{fig:p_con}
\end{figure}

\paragraph{Distribution of stresses}
\begin{figure}[H]
    \centering
    \includegraphics[width=0.9\textwidth]{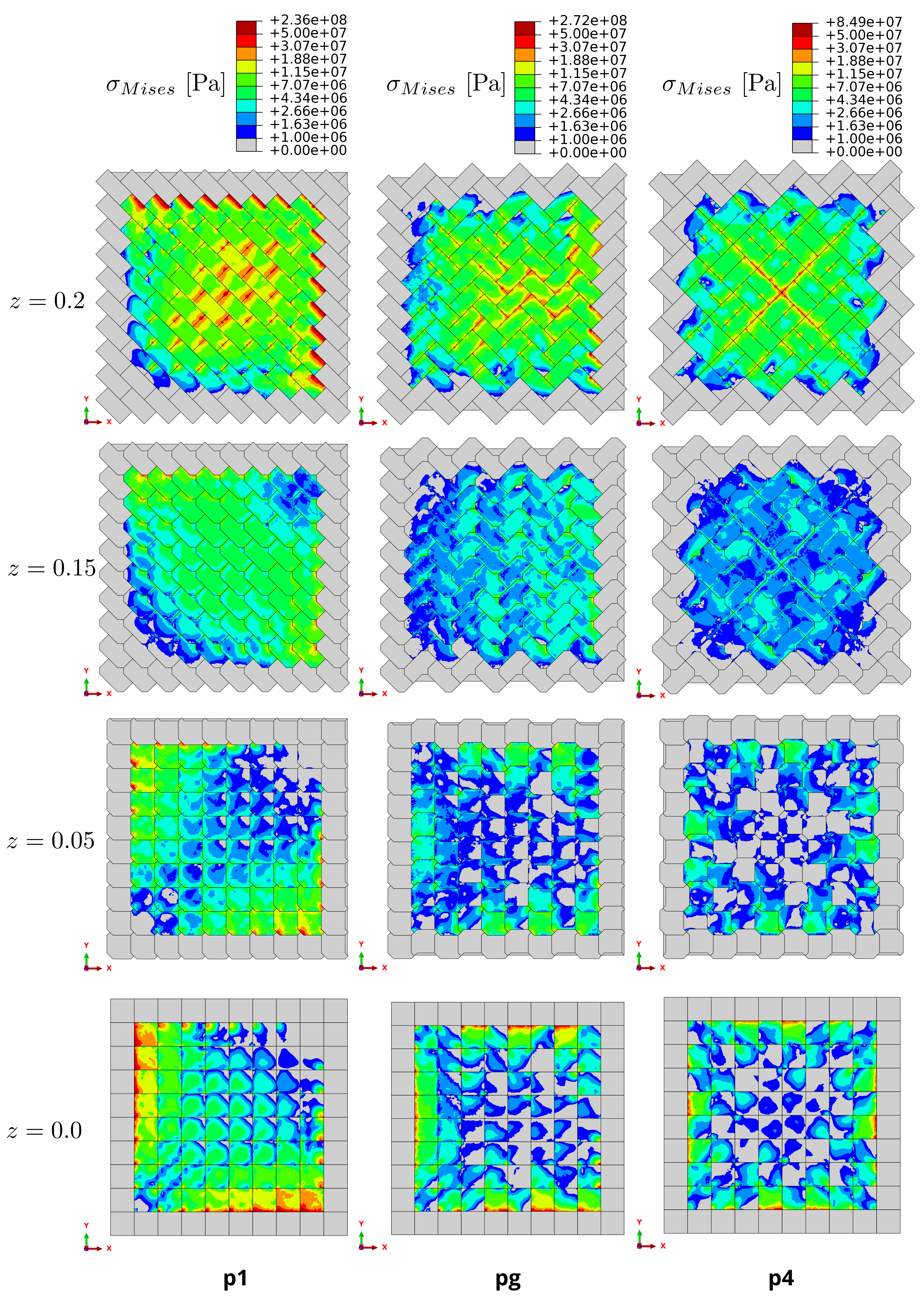}
    \caption{Mises stress distributions in the TIA.}
    \label{fig:Sm}
\end{figure}
\begin{figure}[H]
    \centering
    \includegraphics[width=0.9\textwidth]{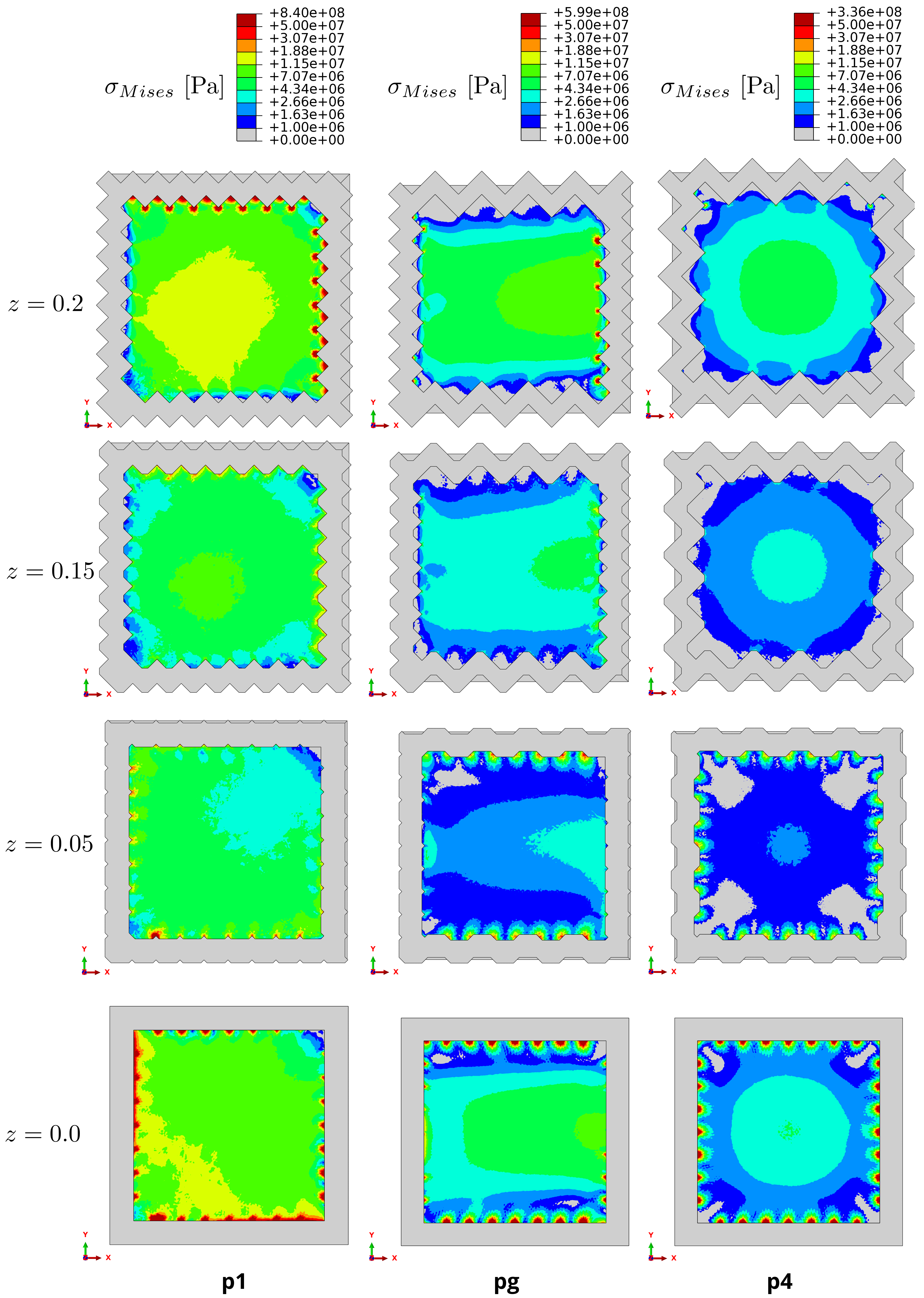}
    \caption{Mises stress distributions in the monolithic plates.}
    \label{fig:Sm_ref}
\end{figure}
Figure \ref{fig:Sm} shows the distribution of Mises stresses in the assemblies. The stress distribution is very complex, which is associated with the non-convex geometry of the block. For comparison, the stress distributions in corresponding solid plates are shown in Figure \ref{fig:Sm_ref}.
Particularly interesting is the distribution of stresses from the top plane, where the load is applied, to the bottom plane. The results show that different block arrangements strongly influence the stress distribution in the assembly. In the top plane ($z=$ \SI{0.2}{\metre}), we can see that the highest stresses occur in the blocks at the top and right boundary as well as centre in $p1$; at the centre and at the right boundary in $pg$; and only in the middle (forming an "X") in the assembly $p4$. Similar observations can  also be made for the reference solutions (Figure \ref{fig:Sm_ref}). Two major differences can be observed between the monolithic plates and the interlocking assemblies. First, the stress distributions in TIA are discontinuous. This observation is not surprising, of course, since the monolithic plate is assumed to be a continuous medium. And second, below the middle plane $z=$ \SI{0.1}{\metre} the majority of the blocks in the assemblies are less stressed than in the monolithic plates, but at the top plane the blocks experience higher stresses. The results also show that where the contact pressure (see Figure \ref{fig:p_con}) between the frame and the assembly is high, the stresses in the blocks are also the highest.
From the results it is also evident that assembly $p4$ distributes stresses in a more optimal way than $pg$ or $p1$ and is therefore the best of the three options for this particular case.
As expected, we observe that the load transfer mechanisms in the regime of compressive stresses are qualitatively similar to those of a monolithic plate, while the load transfer significantly differs in case of tensile stresses (see next paragraph). The reason for this lies in the fact that the gap between neighbouring blocks opens, and thus, contact is lost.
\begin{figure}[H]
    \centering
    \includegraphics[width=1.0\textwidth]{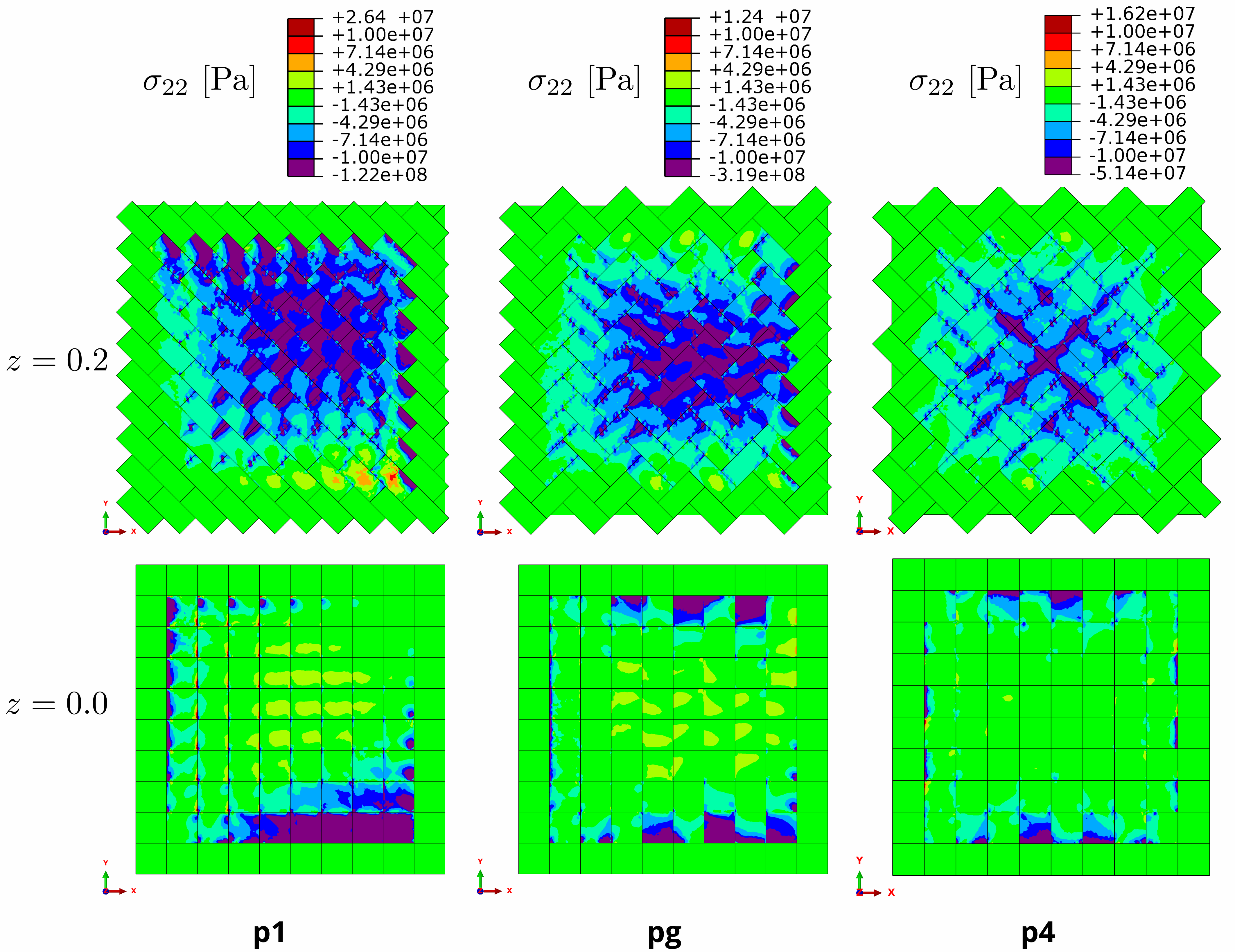}
    \caption{Normal stresses in $y$-direction of the TIA.}
    \label{fig:S_22}
\end{figure}
In civil engineering, the assemblies presented in this paper could be used for example as ceilings. Typically, such ceilings are made of concrete, a material that can withstand high compressive but low tensile loads. A solid plate loaded in the transverse direction will bend, resulting in compression in the upper and tension in the lower half of the plate, which is not desirable. To investigate the behaviour of TIA with respect to this problem, we plotted the normal stress field in $y$ direction (see Figure \ref{fig:S_22}), which is in our case the first indicator of tension or compression in the material. We can see that the negative stress predominates in the top and bottom planes of the TIA. Comparing this result with the corresponding solid plates (Figure \ref{fig:S_22_ref}), we clearly see the advantage of the TIA.

\paragraph{Contact pressure}
One of the most important questions regarding the interlocking mechanism is: How is an external load transferred through the assembly to the bounding frame? To answer this question, we investigate the contact pressure between the blocks and the frame. The results are shown in Figure \ref{fig:p_con}. Since there is no friction involved, contact pressure indirectly represents the distribution of forces in a TIA. Intuitively, one would think that all peripheral blocks would be equally involved in transferring the external load to the frame. The results show (see Figure \ref{fig:p_con}) that this is not the case. Especially, the assemblies $pg$ and $p4$ show a behaviour that deviates from expectation. Furthermore, the simulations show that load transfer of $p1$ occurs over the two boundary edges in the bottom plane and the two opposing boundary edges in the top plane; of $pg$ over three boundary edges in the bottom plane and the opposing one edge in the top plane; and of $p4$ over all boundary edges in the bottom plane only. As the load transfer of $p1$ and $pg$ takes place in two planes, the question arises as to how great the influence of the height of the assembly is on the contact pressure. 
We have two main findings: First, unlike the monolithic plate, the assembly transfers the load mainly by pressure, and thus the resulting contact forces are higher on average. Second, the load is applied to the frame in a patterned manner, whereas the monolithic plate distributes the load continuously along the frame.

\paragraph{Contact pressure on a single block}
\begin{figure}[H]
    \centering
    \includegraphics[width=1.0\textwidth]{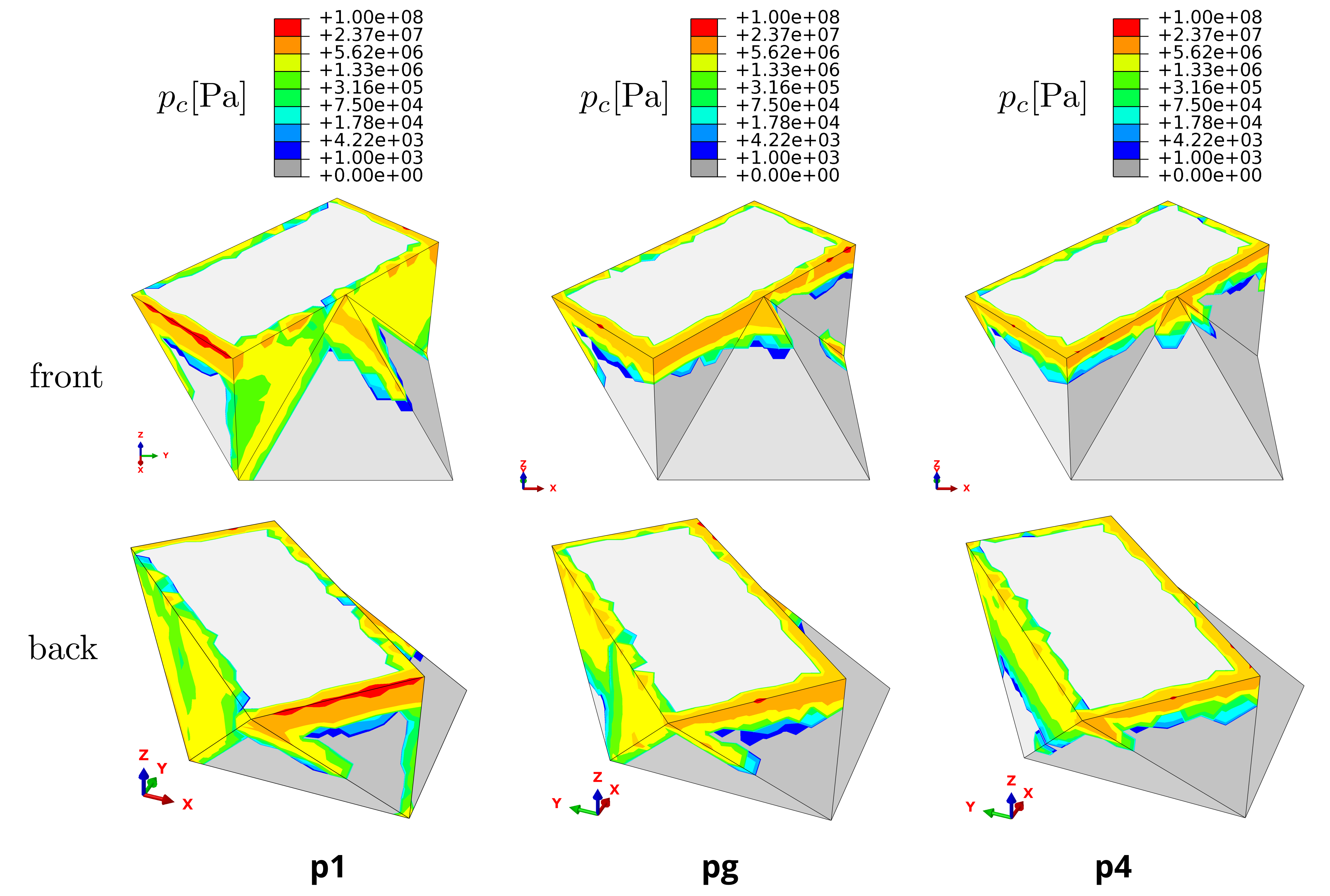}
    \caption{Contact pressure of middle block in the TIA.}
    \label{fig:p_con_block}
\end{figure}
From each TIA, we examined a block near the centre of the assembly to gain insight into which parts of the block are most heavily loaded. For that we simulated the contact pressure present on the surface of the block (Figure \ref{fig:p_con_block}). The central block was chosen because high contact pressures occur around the centre of the assembly for all TIA. For all three blocks, the highest contact pressure appears at the four edges of the rectangle in the top plane. The contact area is quite narrow, which means that this part of the block is subjected to the greatest load. The lower parts of the blocks are practically not loaded at all. Our results suggest that depending on the position of the block within the overall arrangement, the area of contact of the block with its neighbours may vary.  

\newpage

\section{Interlocking Flows} \label{sec:comb-tool}

For applications and fast reliable evaluations, it is necessary to predict the quality of a topological interlocking assembly before its manufacturing or even before running simulations. Since FEM simulations are often time consuming, we introduce a fast discrete criterion based on the combinatorial theory of tilings and flow networks, which we call `Interlocking Flows'. This is essential for identifying candidates of interlocking assemblies that can be further investigated by using more established methods. Here, the focus lies on load transfer from the assembly onto the frame. For this endeavour, we give a combinatorial interpretation of the results presented in Section \ref{sec:mech-inv} and derive a combinatorial model which leads to a stable and fast method for giving a first heuristic of the mechanical performance of an interlocking assembly.

One discrete tool to study topological interlocking assemblies are \emph{Directional Blocking Graphs}, which are introduced in \cite{WilsonPhDThesis,WilsonLatombe94} and investigated in the context of interlocking assemblies in \cite{wang_desia_2018}. Here, we give an adapted version of the definition of such a graph for interlocking assemblies by treating the blocks on the frame differently. For this let $A=\{ X_i \mid i\in I\}$, be a topological interlocking assembly consisting of blocks $X_i\subset \R^3$ indexed by a finite index set $I$ with a frame indexed by $J\subset I$, the core indexed by $C \coloneqq I \setminus J$ and $d\in \R^3$ a vector. We say that a block in the core is restrained in direction $d$ by another block if shifting the first block in the direction $d$ leads to an intersection with the latter block, i.e. for $i\in C$ and $j\in I$, the translated block $X_i-d\coloneqq \{x-d \mid x\in X_i \}$ intersects with $X_j$. Furthermore, we say that a block in the frame restrains itself from moving.

\begin{definition}
 The \emph{Directional Blocking Graph} (short \emph{DGB}) $\mathcal{G}(A,d)$ is defined as the directed graph with
\begin{enumerate}
    \item vertices given by the set $I$ and
    \item arcs of the form $i\to j$ if the block $X_i$ is restrained by $X_j$ in direction $d$ for $i,j\in I$.
\end{enumerate}   
\end{definition}

\begin{figure}[H]
    \centering
    \begin{subfigure}[b]{0.49\textwidth}
        \centering
        \includegraphics[height=4cm]{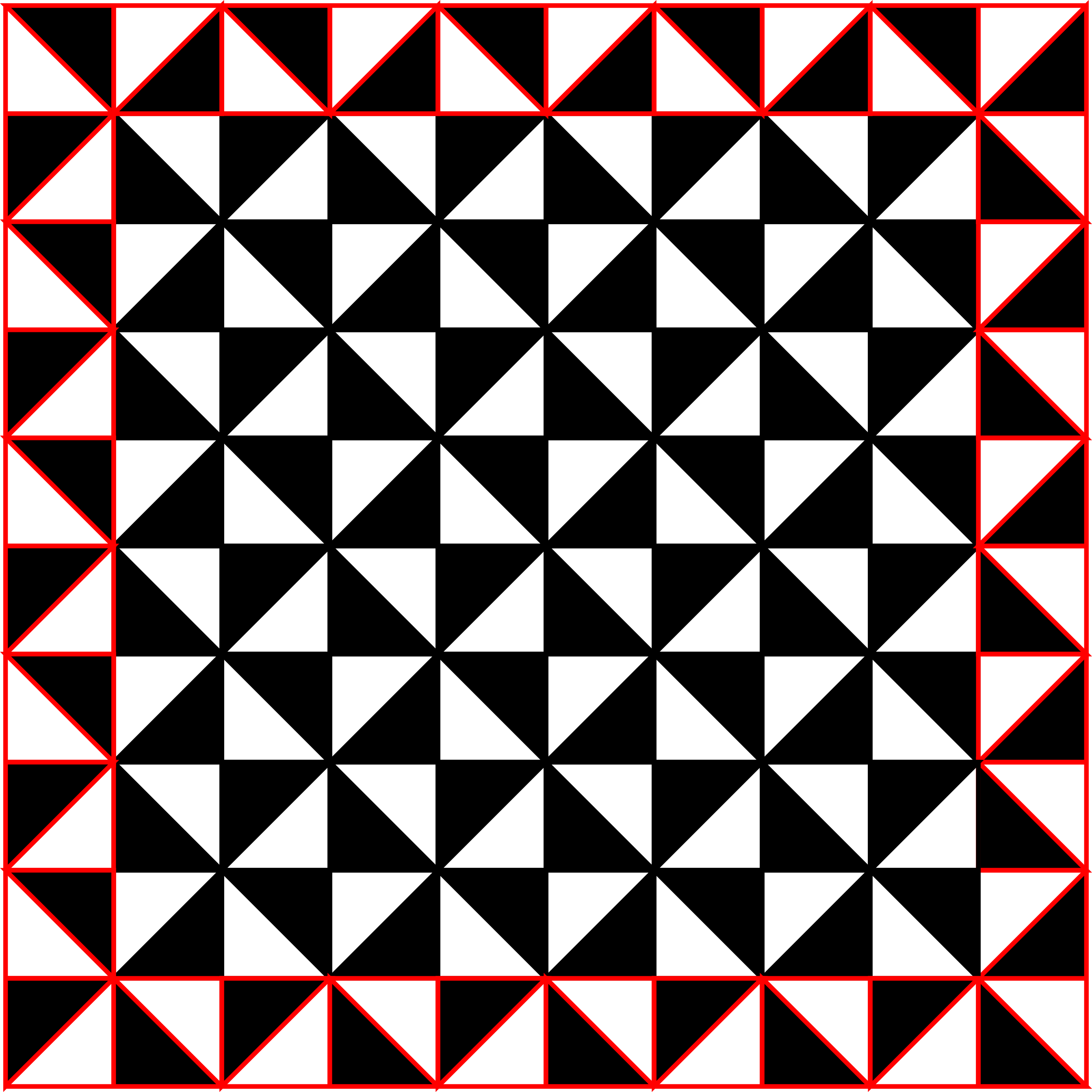}
        \caption{$p4$ Truchet tiling}    
        \label{fig:p4_truchet}
    \end{subfigure}
     \begin{subfigure}[b]{0.49\textwidth}
        \centering
        \includegraphics[height=4cm]{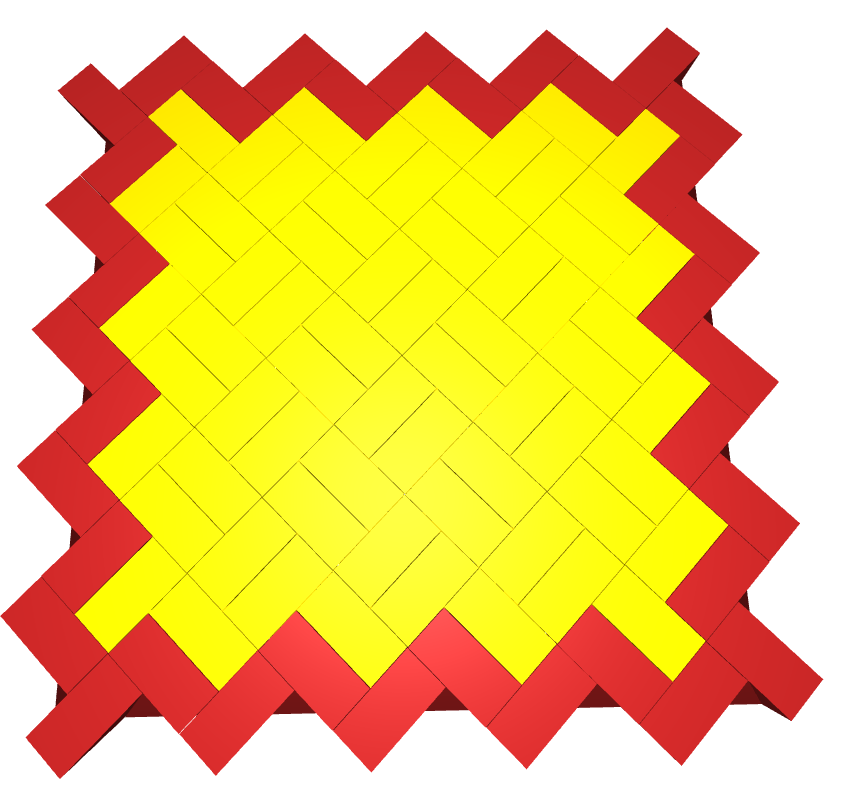}
        \caption{$p4$ assembly with frame}   
        \label{fig:p4_frame}
    \end{subfigure}
     \begin{subfigure}[b]{0.49\textwidth}
        \centering
        \includegraphics[height=4cm]{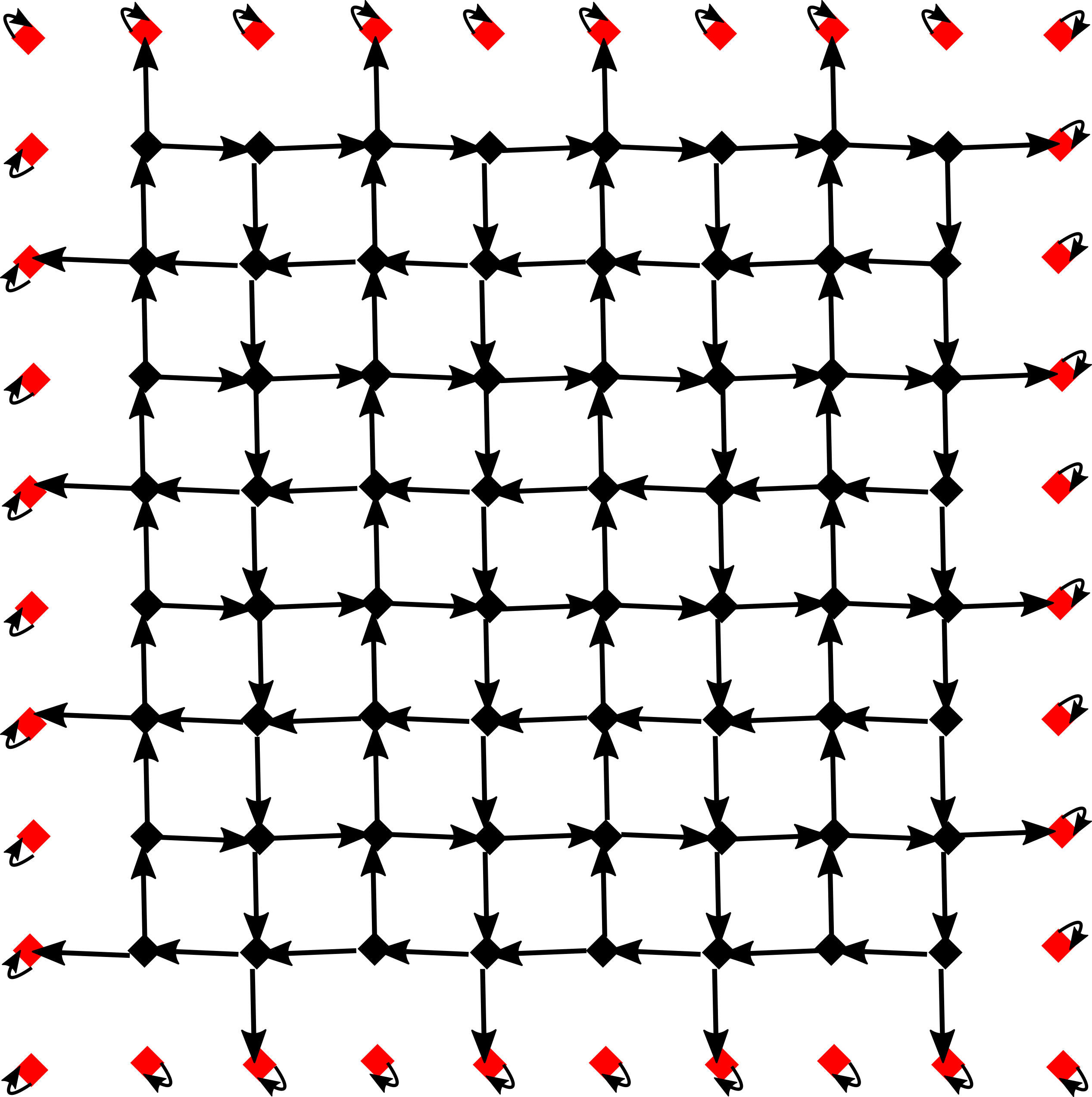}
        \caption{$p4$ DGB}  
        \label{fig:p4_network}
    \end{subfigure}
    \begin{subfigure}[b]{0.49\textwidth}
        \centering
        \resizebox{4cm}{!}{\begin{tikzpicture}
  \matrix (m) [matrix of math nodes,
    nodes in empty cells,
    row sep=-\pgflinewidth,
    column sep=-\pgflinewidth,
    nodes={draw, minimum size=1cm, anchor=center, inner sep=0pt, font=\large}
  ] {
    |[red]|0 & |[red]|0 & |[red]|0 & |[red]|0 & |[red]|0 & |[red]|0 & |[red]|0 & |[red]|0 & |[red]|0 & |[red]|0 \\
    |[red]|0 & |[black]|1 & |[black]|1 & |[black]|1 & |[black]|1 & |[black]|1 & |[black]|1 & |[black]|1 & |[black]|1 & |[red]|0 \\
   |[red]|0 & |[black]|1 & |[black]|1 & |[black]|1 & |[black]|1 & |[black]|1 & |[black]|1 & |[black]|1 & |[black]|1 & |[red]|0 \\
   |[red]|0 & |[black]|1 & |[black]|1 & |[black]|1 & |[black]|1 & |[black]|1 & |[black]|1 & |[black]|1 & |[black]|1 & |[red]|0 \\
    |[red]|0 & |[black]|1 & |[black]|1 & |[black]|1 & |[black]|1 & |[black]|1 & |[black]|1 & |[black]|1 & |[black]|1 & |[red]|0 \\
    |[red]|0 & |[black]|1 & |[black]|1 & |[black]|1 & |[black]|1 & |[black]|1 & |[black]|1 & |[black]|1 & |[black]|1 & |[red]|0 \\
    |[red]|0 & |[black]|1 & |[black]|1 & |[black]|1 & |[black]|1 & |[black]|1 & |[black]|1 & |[black]|1 & |[black]|1 & |[red]|0 \\
    |[red]|0 & |[black]|1 & |[black]|1 & |[black]|1 & |[black]|1 & |[black]|1 & |[black]|1 & |[black]|1 & |[black]|1 & |[red]|0 \\
    |[red]|0 & |[black]|1 & |[black]|1 & |[black]|1 & |[black]|1 & |[black]|1 & |[black]|1 & |[black]|1 & |[black]|1 & |[red]|0 \\
    |[red]|0 & |[red]|0 & |[red]|0 & |[red]|0 & |[red]|0 & |[red]|0 & |[red]|0 & |[red]|0 & |[red]|0 & |[red]|0 \\
  };
\end{tikzpicture}}
        \caption{$x\in \mathbb{R}^I\cong \mathbb{R}^{100}\cong \mathbb{R}^{10 \times 10}$}  
        \label{fig:p4_truchet_0}
    \end{subfigure}
    \begin{subfigure}[b]{0.49\textwidth}
        \centering
        \resizebox{4cm}{!}{\begin{tikzpicture}
  \matrix (m) [matrix of math nodes,
    nodes in empty cells,
    row sep=-\pgflinewidth,
    column sep=-\pgflinewidth,
    nodes={draw, minimum size=1cm, anchor=center, inner sep=0pt, font=\large}
  ] {
    |[red]|0 & |[red]|\frac{1}{2} & |[red]|0 & |[red]|\frac{1}{2} & |[red]|0 & |[red]|\frac{1}{2} & |[red]|0 & |[red]|\frac{1}{2} & |[red]|0 & |[red]|0 \\
    |[red]|0 & |[black]|\frac{1}{2} & |[black]|\frac{1}{2} & |[black]|1 & |[black]|\frac{1}{2} & |[black]|1 & |[black]|\frac{1}{2} & |[black]|1 & |[black]|\frac{1}{2} & |[red]|\frac{1}{2} \\
    |[red]|\frac{1}{2} & |[black]|1 & |[black]|1 & |[black]|1 & |[black]|1 & |[black]|1 & |[black]|1 & |[black]|1 & |[black]|\frac{1}{2} & |[red]|0 \\
    |[red]|0 & |[black]|\frac{1}{2} & |[black]|1 & |[black]|1 & |[black]|1 & |[black]|1 & |[black]|1 & |[black]|1 & |[black]|1 & |[red]|\frac{1}{2} \\
    |[red]|\frac{1}{2} & |[black]|1 & |[black]|1 & |[black]|1 & |[black]|1 & |[black]|1 & |[black]|1 & |[black]|1 & |[black]|\frac{1}{2} & |[red]|0 \\
    |[red]|0 & |[black]|\frac{1}{2} & |[black]|1 & |[black]|1 & |[black]|1 & |[black]|1 & |[black]|1 & |[black]|1 & |[black]|1 & |[red]|\frac{1}{2} \\
    |[red]|\frac{1}{2} & |[black]|1 & |[black]|1 & |[black]|1 & |[black]|1 & |[black]|1 & |[black]|1 & |[black]|1 & |[black]|\frac{1}{2} & |[red]|0 \\
    |[red]|0 & |[black]|\frac{1}{2} & |[black]|1 & |[black]|1 & |[black]|1 & |[black]|1 & |[black]|1 & |[black]|1 & |[black]|1 & |[red]|\frac{1}{2} \\
    |[red]|\frac{1}{2} & |[black]|\frac{1}{2} & |[black]|1 & |[black]|\frac{1}{2} & |[black]|1 & |[black]|\frac{1}{2} & |[black]|1 & |[black]|\frac{1}{2} & |[black]|\frac{1}{2} & |[red]|0 \\
    |[red]|0 & |[red]|0 & |[red]|\frac{1}{2} & |[red]|0 & |[red]|\frac{1}{2} & |[red]|0 & |[red]|\frac{1}{2} & |[red]|0 & |[red]|\frac{1}{2} & |[red]|0 \\
  };
\end{tikzpicture}}
        \caption{$A\cdot x$}   
        \label{fig:p4_truchet_1}
    \end{subfigure}
    \begin{subfigure}[b]{0.49\textwidth}
        \centering
        \resizebox{4cm}{!}{\begin{tikzpicture}
  \matrix (m) [matrix of math nodes,
    nodes in empty cells,
    row sep=-\pgflinewidth,
    column sep=-\pgflinewidth,
    nodes={draw, minimum size=1cm, anchor=center, inner sep=0pt, font=\large}
  ] {
    |[red]|0 & |[red]|2.58 & |[red]|0 & |[red]|4.38 & |[red]|0 & |[red]|4.88 & |[red]|0 & |[red]|4.16 & |[red]|0 & |[red]|0 \\
    |[red]|0 & |[black]|0 & |[black]|0 & |[black]|0 & |[black]|0 & |[black]|0 & |[black]|0 & |[black]|0 & |[black]|0 & |[red]|2.58 \\
    |[red]|4.16 & |[black]|0 & |[black]|0 & |[black]|0 & |[black]|0 & |[black]|0 & |[black]|0 & |[black]|0 & |[black]|0 & |[red]|0 \\
    |[red]|0 & |[black]|0 & |[black]|0 & |[black]|0 & |[black]|0 & |[black]|0 & |[black]|0 & |[black]|0 & |[black]|0 & |[red]|4.38 \\
    |[red]|4.88 & |[black]|0 & |[black]|0 & |[black]|0 & |[black]|0 & |[black]|0 & |[black]|0 & |[black]|0 & |[black]|0 & |[red]|0 \\
    |[red]|0 & |[black]|0 & |[black]|0 & |[black]|0 & |[black]|0 & |[black]|0 & |[black]|0 & |[black]|0 & |[black]|0 & |[red]|4.88 \\
    |[red]|4.38 & |[black]|0 & |[black]|0 & |[black]|0 & |[black]|0 & |[black]|0 & |[black]|0 & |[black]|0 & |[black]|0 & |[red]|0 \\
    |[red]|0 & |[black]|0 & |[black]|0 & |[black]|0 & |[black]|0 & |[black]|0 & |[black]|0 & |[black]|0 & |[black]|0 & |[red]|4.16 \\
    |[red]|2.58 & |[black]|0 & |[black]|0 & |[black]|0 & |[black]|0 & |[black]|0 & |[black]|0 & |[black]|0 & |[black]|0 & |[red]|0 \\
    |[red]|0 & |[red]|0 & |[red]|4.16 & |[red]|0 & |[red]|4.88 & |[red]|0 & |[red]|4.38 & |[red]|0 & |[red]|2.58 & |[red]|0 \\
  };
\end{tikzpicture}}
        \caption{$A^n\cdot x$ for $n$ large rounded to two digits}  
        \label{fig:p4_truchet_n}
    \end{subfigure}
    \caption{Combinatorial Interpretation of $p4$ experiments}
    \label{fig:p4_discrete}
\end{figure}

The planar assemblies of copies of the Versatile Block are characterised by Truchet tiles, see \cite{bridges23}. Here a single oriented Truchet tile corresponds to a single oriented Versatile Block. The admissible assemblies are then characterised by the rule that two Truchet tiles only meet at alternating colours. For example, Figure \ref{fig:p4_frame} displays the $p4$ assembly $A_{p4}$ with $|I|=100$ where the frame is marked in red. The corresponding Truchet tiling is given in Figure \ref{fig:p4_truchet}.

\begin{figure}[H]
    \centering
    \begin{subfigure}[b]{0.49\textwidth}
        \centering
        \includegraphics[height=4cm]{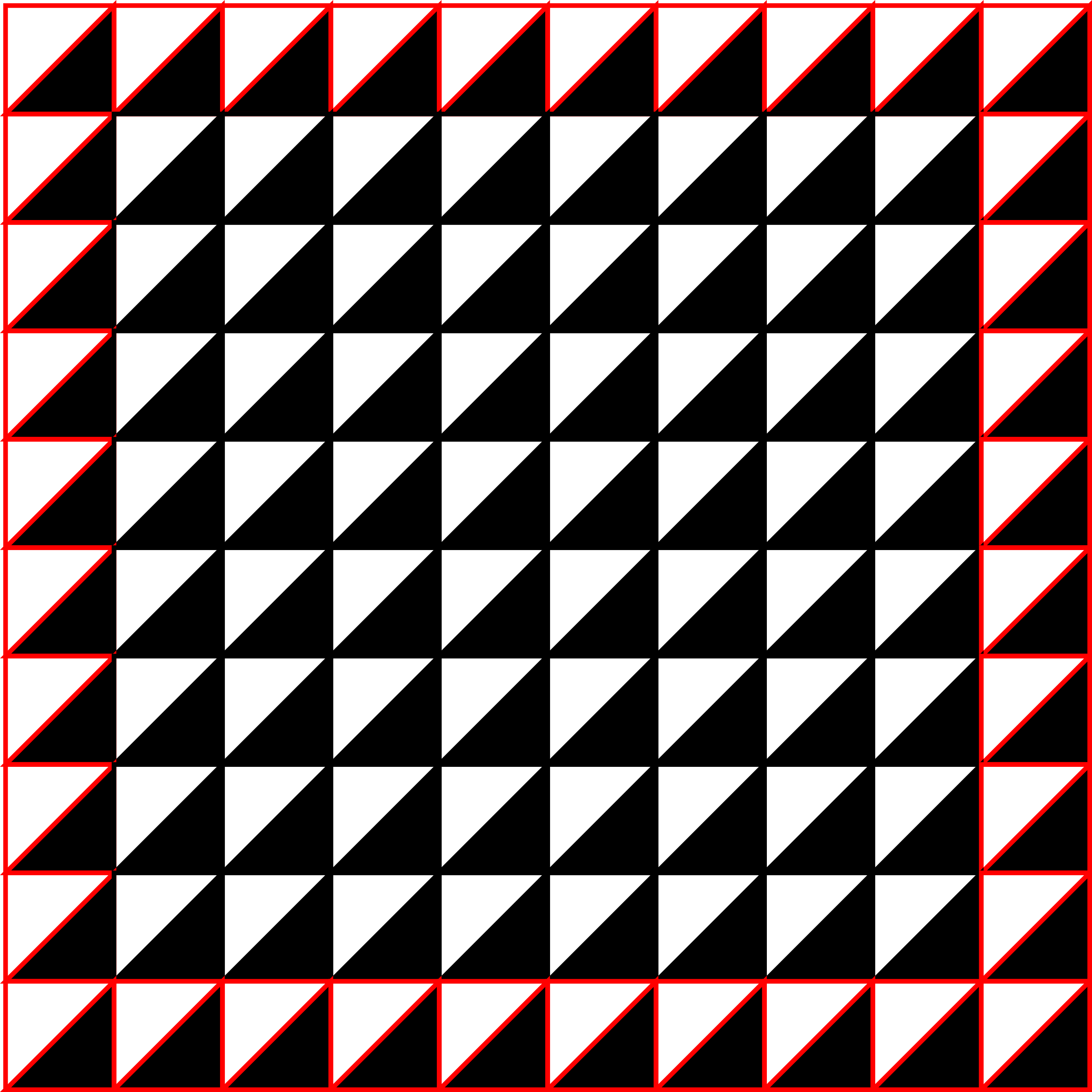}
        \caption{$p1$ Truchet tiling}    
        \label{fig:p1_truchet}
    \end{subfigure}
     \begin{subfigure}[b]{0.49\textwidth}
        \centering
        \includegraphics[height=4cm]{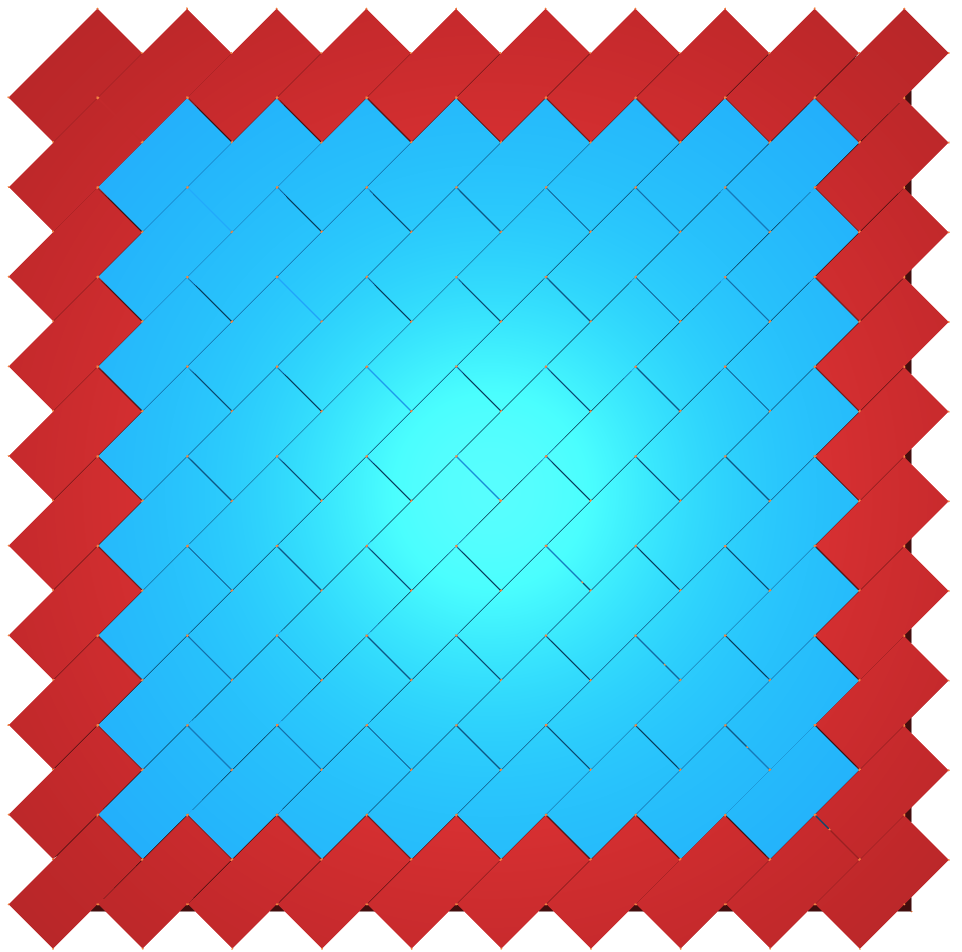}
        \caption{$p1$ assembly with frame}   
        \label{fig:p1_frame}
    \end{subfigure}
     \begin{subfigure}[b]{0.49\textwidth}
        \centering
        \includegraphics[height=4cm]{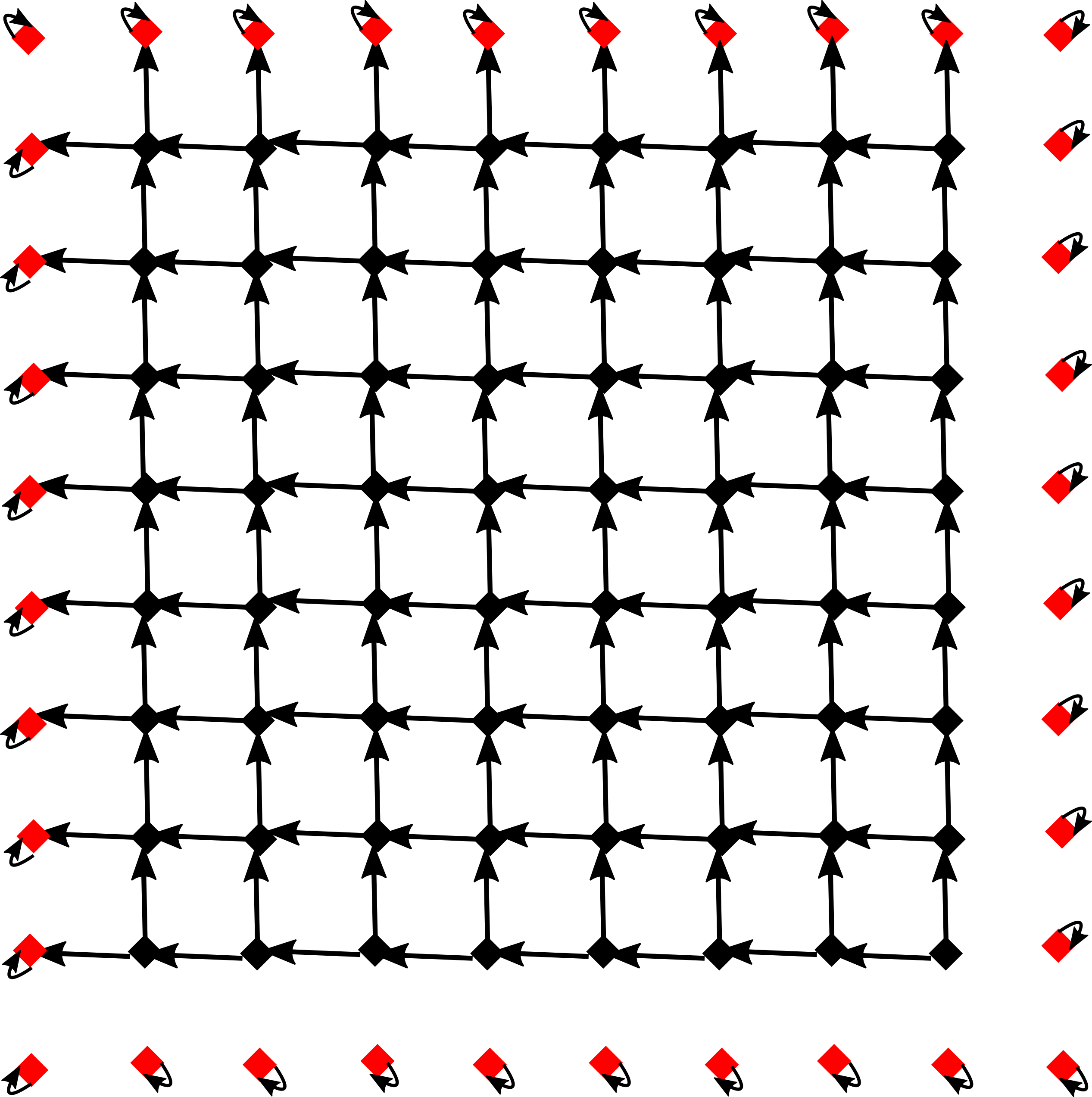}
        \caption{$p1$ DGB}  
        \label{fig:p1_network}
    \end{subfigure}
    \begin{subfigure}[b]{0.49\textwidth}
        \centering
        \resizebox{4cm}{!}{\begin{tikzpicture}
  \matrix (m) [matrix of math nodes,
    nodes in empty cells,
    row sep=-\pgflinewidth,
    column sep=-\pgflinewidth,
    nodes={draw, minimum size=1cm, anchor=center, inner sep=0pt, font=\large}
  ] {
    |[red]|0 & |[red]|6.43 & |[red]|5.93 & |[red]|5.32 & |[red]|4.61 & |[red]|3.81 & |[red]|2.92 & |[red]|1.98 & |[red]|1.00 & |[red]|0 \\
    |[red]|6.43 & |[black]|0 & |[black]|0 & |[black]|0 & |[black]|0 & |[black]|0 & |[black]|0 & |[black]|0 & |[black]|0 & |[red]|0 \\
    |[red]|5.93 & |[black]|0 & |[black]|0 & |[black]|0 & |[black]|0 & |[black]|0 & |[black]|0 & |[black]|0 & |[black]|0 & |[red]|0 \\
    |[red]|5.32 & |[black]|0 & |[black]|0 & |[black]|0 & |[black]|0 & |[black]|0 & |[black]|0 & |[black]|0 & |[black]|0 & |[red]|0 \\
    |[red]|4.61 & |[black]|0 & |[black]|0 & |[black]|0 & |[black]|0 & |[black]|0 & |[black]|0 & |[black]|0 & |[black]|0 & |[red]|0 \\
    |[red]|3.81 & |[black]|0 & |[black]|0 & |[black]|0 & |[black]|0 & |[black]|0 & |[black]|0 & |[black]|0 & |[black]|0 & |[red]|0 \\
    |[red]|2.92 & |[black]|0 & |[black]|0 & |[black]|0 & |[black]|0 & |[black]|0 & |[black]|0 & |[black]|0 & |[black]|0 & |[red]|0 \\
    |[red]|1.98 & |[black]|0 & |[black]|0 & |[black]|0 & |[black]|0 & |[black]|0 & |[black]|0 & |[black]|0 & |[black]|0 & |[red]|0 \\
    |[red]|1.00 & |[black]|0 & |[black]|0 & |[black]|0 & |[black]|0 & |[black]|0 & |[black]|0 & |[black]|0 & |[black]|0 & |[red]|0 \\
    |[red]|0 & |[red]|0 & |[red]|0 & |[red]|0 & |[red]|0 & |[red]|0 & |[red]|0 & |[red]|0 & |[red]|0 & |[red]|0 \\
  };
\end{tikzpicture}}
        \caption{ $A^n\cdot x$ for $n$ large rounded to two digits}  
        \label{fig:p1_truchet_n}
    \end{subfigure}
    \caption{Combinatorial Interpretation of $p1$ experiments}
    \label{fig:p1_discrete}
\end{figure}

\begin{figure}[H]
    \centering
    \begin{subfigure}[b]{0.49\textwidth}
        \centering
        \includegraphics[height=4cm]{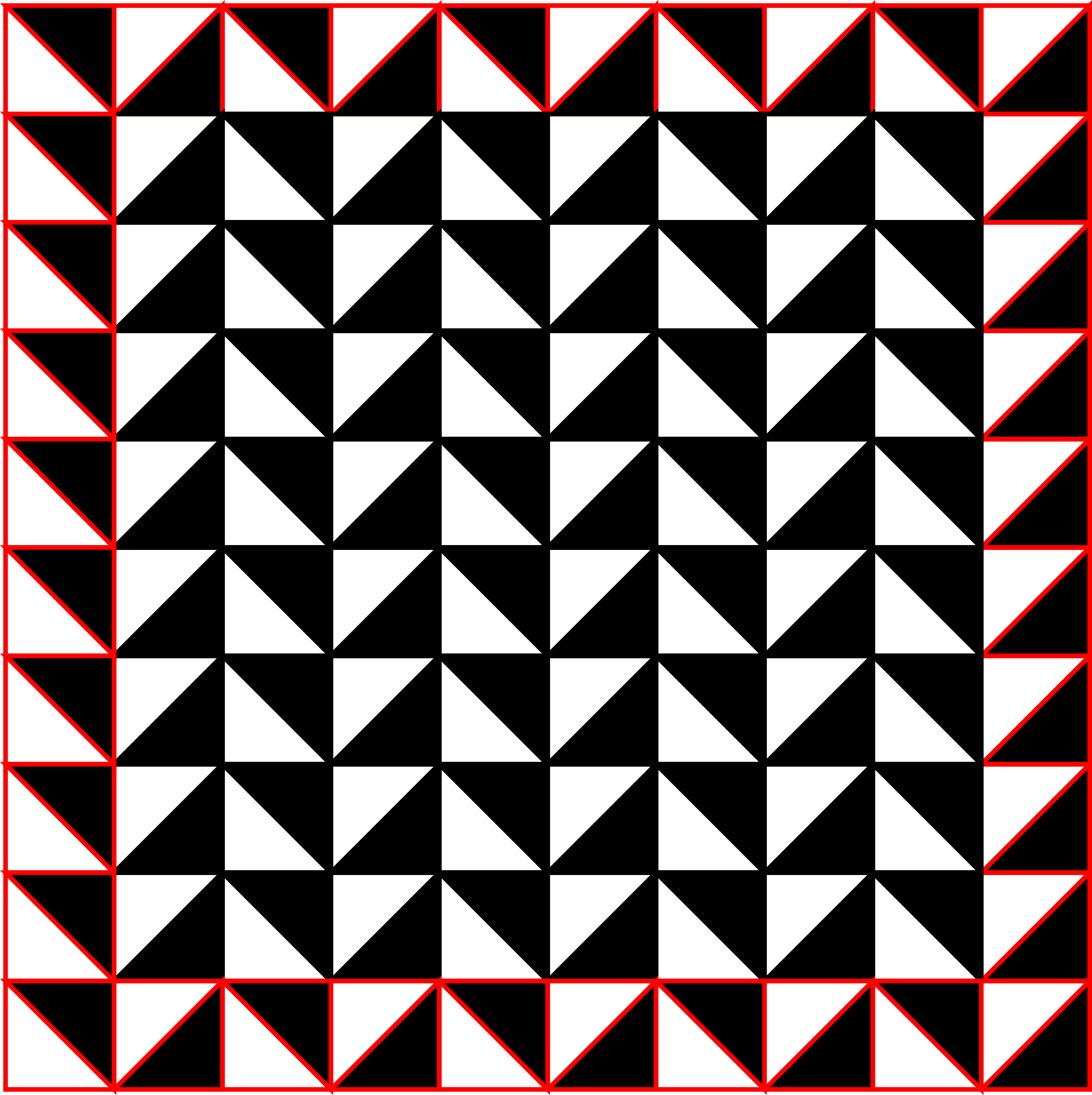}
        \caption{$pg$ Truchet tiling}    
        \label{fig:pg_truchet}
    \end{subfigure}
     \begin{subfigure}[b]{0.49\textwidth}
        \centering
        \includegraphics[height=4cm]{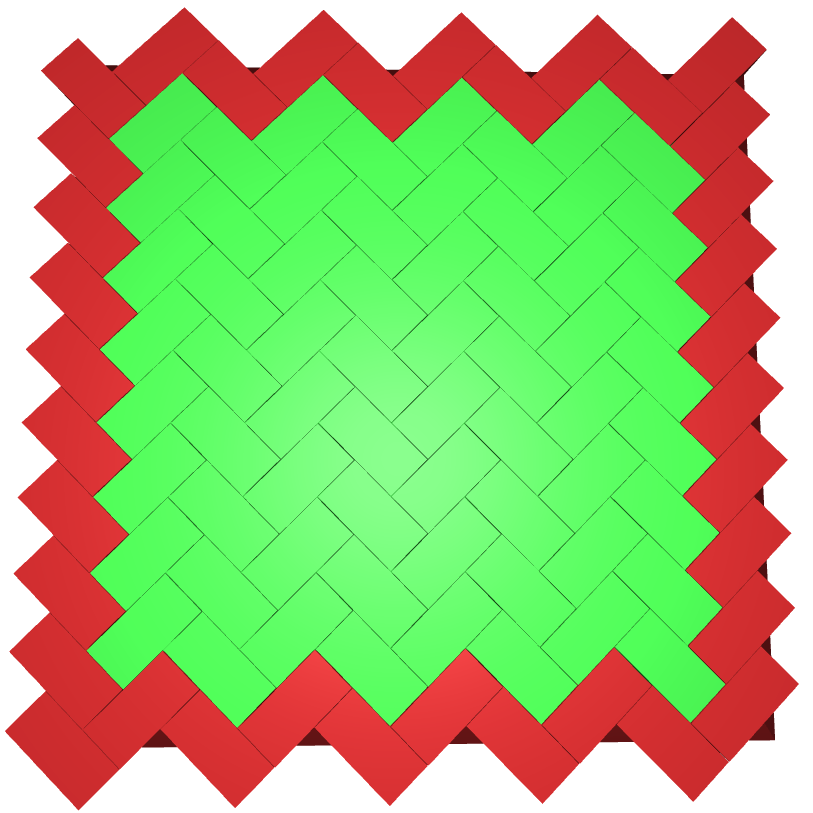}
        \caption{$pg$ assembly with frame}   
        \label{fig:pg_frame}
    \end{subfigure}
     \begin{subfigure}[b]{0.49\textwidth}
        \centering
        \includegraphics[height=4cm]{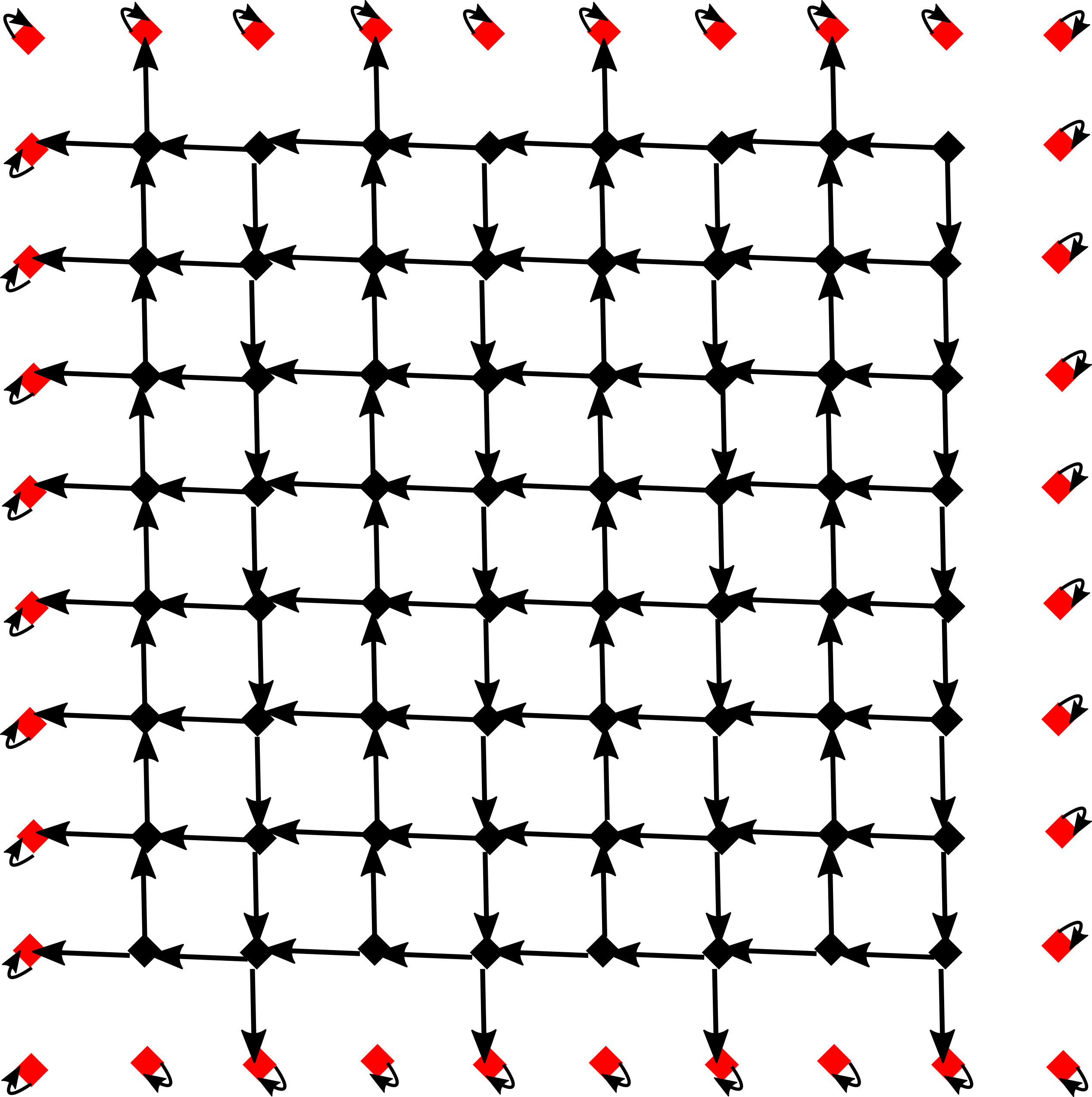}
        \caption{$pg$ DGB}  
        \label{fig:pg_network}
    \end{subfigure}
    \begin{subfigure}[b]{0.49\textwidth}
        \centering
        \resizebox{4cm}{!}{\begin{tikzpicture}
  \matrix (m) [matrix of math nodes,
    nodes in empty cells,
    row sep=-\pgflinewidth,
    column sep=-\pgflinewidth,
    nodes={draw, minimum size=1cm, anchor=center, inner sep=0pt, font=\large}
  ] {
    |[red]|0 & |[red]|4.44 & |[red]|0 & |[red]|3.71 & |[red]|0 & |[red]|2.82 & |[red]|0 & |[red]|1.66 & |[red]|0 & |[red]|0 \\
    |[red]|4.44 & |[black]|0 & |[black]|0 & |[black]|0 & |[black]|0 & |[black]|0 & |[black]|0 & |[black]|0 & |[black]|0 & |[red]|0 \\
    |[red]|5.52 & |[black]|0 & |[black]|0 & |[black]|0 & |[black]|0 & |[black]|0 & |[black]|0 & |[black]|0 & |[black]|0 & |[red]|0 \\
    |[red]|6.10 & |[black]|0 & |[black]|0 & |[black]|0 & |[black]|0 & |[black]|0 & |[black]|0 & |[black]|0 & |[black]|0 & |[red]|0 \\
    |[red]|6.27 & |[black]|0 & |[black]|0 & |[black]|0 & |[black]|0 & |[black]|0 & |[black]|0 & |[black]|0 & |[black]|0 & |[red]|0 \\
    |[red]|6.05 & |[black]|0 & |[black]|0 & |[black]|0 & |[black]|0 & |[black]|0 & |[black]|0 & |[black]|0 & |[black]|0 & |[red]|0 \\
    |[red]|5.42 & |[black]|0 & |[black]|0 & |[black]|0 & |[black]|0 & |[black]|0 & |[black]|0 & |[black]|0 & |[black]|0 & |[red]|0 \\
    |[red]|4.30 & |[black]|0 & |[black]|0 & |[black]|0 & |[black]|0 & |[black]|0 & |[black]|0 & |[black]|0 & |[black]|0 & |[red]|0 \\
    |[red]|2.55 & |[black]|0 & |[black]|0 & |[black]|0 & |[black]|0 & |[black]|0 & |[black]|0 & |[black]|0 & |[black]|0 & |[red]|0 \\
    |[red]|0 & |[red]|0 & |[red]|4.10 & |[red]|0 & |[red]|3.31 & |[red]|0 & |[red]|2.31 & |[red]|0 & |[red]|1.00 & |[red]|0 \\
  };
\end{tikzpicture}}
        \caption{$A^n\cdot x$ for $n$ large rounded to two digits}  
        \label{fig:pg_truchet_n}
    \end{subfigure}
    \caption{Combinatorial Interpretation of $pg$ experiments}
    \label{fig:pg_discrete}
\end{figure}

In the context of this paper, we apply a load $x$ in the direction $d=(0,0,-\varepsilon)$ for a small value $\varepsilon>0$ for each block in the interlocking assembly simultaneously.
We use the DGB $\mathcal{G}=\mathcal{G}(A_{p4},d)$ to model how this load is transferred onto the frame of the underlying assembly.
This can be achieved by introducing a value function on the arcs of $\mathcal{G}$ yielding a flow network with sinks given by the nodes corresponding to the blocks belonging to the frame.
For the assemblies based on the Versatile Block, we give a value function $v$ for arcs of the DGB $\mathcal{G}$ as follows: let $i\to j$ be an arc of $\mathcal{G}$ with $i,j\in I$, then we set
$$v(i\to j) \coloneqq 
\begin{cases}
    \frac{1}{2}, & \text{if } i\neq j\\
    1,           & \text{if } i=j.
\end{cases}$$
We choose the value $\frac{1}{2}$ for distinct blocks $i,j$ since a given Versatile Block $i$ is supported equally by two of its neighbouring blocks. For blocks $j\in J$ belonging to the frame, we set $v(j\to j)=1$, since these blocks are stabilised and fixed from moving. We extend the value function $v$ to any arcs $i\to j$ not contained in $\mathcal{G}$ by setting $v(i \to j)=0.$
With this choice of value function $v$ it follows that $\sum_{j=1}^{100}v(i\to j)=1$, for all $i\in I$ and together with the fact that all values of $v$ are non-negative, we obtain the (right) \emph{stochastic matrix} $$A=(v(i\to j))_{i,j\in I}\in \R_{\geq 0}^{100\times 100},$$ i.e. the entries of each row of $A$ sum up to $1$. This matrix can be viewed as a weighted \emph{adjacency matrix} of $\mathcal{G}$ yielding a flow network with capacity function $v$.
This results in the following combinatorial interpretation of the experiments presented above: let $x\in \R_{\geq 0}^I$ be a load vector with $x_j=0$,
if $j\in J$ and $x_i\in \R_{\geq 0}$ for $i\in C$. Thus the entries of $x$ correspond to the applied loads in direction $d$ on each block. In the experiments above $x$ is chosen to be $x_i=1$ for $i\in C$ and $x_j=0$ for $j\in J$. The model of load transfer can then be discretised by considering the matrix-vector multiplication $$A^k\cdot x$$ for discrete time steps $k=0,\dots,n$, where $n\gg k$ is chosen to be large with $A^n\cdot x$ being close to the convergence load transfer on the frame given the initial load $x$. Since $A$ is a stochastic matrix, it follows that the sum over all entries of $A^k\cdot x$ equals the sum over the entries of $x$, i.e. $$\sum_{i=1}^{100}(A^k\cdot x)_i= \sum_{i=1}^{100}x_i,$$ which can be interpreted as a discrete version of a conservation of energy law. The matrix vector multiplication $A^k\cdot x$ can be computed by exploiting the flow structure as follows: 
\begin{enumerate}
    \item create an empty square grid corresponding to the Truchet tiling
    \item fill the  box corresponding $i\in I$ with the value $x_i$ (initialize vector $x$)
    \item add $1/2$ times the value of box $i\in C$ to box $j$ if the white part of box $i$ touches the black part of box $j$ (this corresponds to the matrix multiplication $A\cdot x)$
    \item iterate the second step $k-1$ times with the updated boxes.
\end{enumerate}

\paragraph{Comparison to FEM Analysis}
We note that the modelled  load transfer onto the frame agrees with the results obtained by the FEM analysis (compare Figure \ref{fig:p_con} and Figure \ref{fig:p4_discrete}-\ref{fig:pg_discrete}). This leads to a fast discrete evaluation criterion which is of interest as there are many assemblies using the Versatile Block. It yields a fast impression of how forces transfer onto the frame. In this way, we can pick candidates for certain applications in a more time-efficient manner. For an $8\times 8$ interlocking assembly with $64$ for blocks, there are $2^{8+8-3}=2^{13}=8192$ possible assemblies (up to rotation and mirroring) using the Versatile Block, see \cite{bridges23}. For an initial prediction, we can evaluate the force transfer onto the frame for all $8192$ assemblies in a matter of seconds as it only revolves around matrix-vector multiplications with relatively small quadratic matrices, i.e. $100\times 100$. This is a major advantage for gaining an initial response as FEM simulations are in general not as cost- and time-efficient. For future research, this discrete evaluation criterion needs to be further investigated and refined to predict how the load transfer from the blocks onto the frame occurs.

\section{Discussion}

In this work we focused on topological interlocking assemblies consisting of homogeneous and isotropic blocks. We applied assembly strategies based on wallpaper symmetries and used a fixed shape of each block.
\subsection{Future work}
\paragraph{Influence of geometric properties}
The authors of \cite{bridges23} have shown that there are exponentially many possibilities to arrange the Versatile Block into a topological interlocking assembly. Investigating several assemblies which do not admit a wallpaper symmetry might extend the understanding of the influence of arrangement on the mechanical performance of topological interlocking assemblies.

The blocks investigated in this paper were created using wallpaper groups. Using the techniques in \cite{topological22} it is possible to create many different blocks and the influence of the choice of block on the mechanical performance can be investigated.

It is well known that a structure under load experiences stress spikes in its corners. It needs to be investigated if the design of the block can be modified to reduce these spikes. 

Another factor influencing the performance of the topological interlocking assembly might be the distance between the top and bottom plane of the planar assembly. In order to design material minimised components the mechanical performance of relatively thin assemblies should be investigated.

\paragraph{Influence of mechanical properties}
In the simulations discussed here friction has been ignored to isolate the effect the arrangement has on the mechanical performance. Moreover, the force has always been applied to the top plane. 
In general, the impact of friction and applying the load to the bottom plane should be investigated.

To gain an insight into how the assembly reacts to loads in general, the load is applied evenly to all rectangles (each consisting of three triangular faces) on the top plane. However, in certain applications of such an assembly, the loads could be more localised. To understand how the assembly reacts to more localised loads and especially how the stresses are transferred throughout the assembly, additional simulations are required. Gaining an insight into this stress distribution could also guide the placement and choice of reinforcement inside each block.

One of the main goals of using reinforced materials is to design resource-efficient and lightweight components. 
Therefore, it is paramount to test whether interlocking assemblies constructed by removing the interior of all blocks display a similar performance properties to assemblies constructed of solid blocks. 
Note that the interior of hollow blocks could be filled  with a different material to influence other properties, e.g.\ by inserting insulating material, sound proofing material or using the space for electrical wiring.

\subsection{Conclusion}
The arrangement of the blocks in topological interlocking assemblies has a significant influence on the structural behaviour of the overall assembly such as its point of deflection, the load transfer mechanisms, as well as stress distribution. We investigated three different symmetric arrangements of the Versatile Block and showcased these differences in mechanical behaviour by FEM analyses. Moreover, by comparison with a monolithic plate we demonstrated that the structural behaviour of TIA is qualitatively and quantitatively in the same order of magnitude. Further, we developed a combinatorial tool that is capable of pre-evaluating the load transfer onto the frame. In summary we showed that by exploiting the rich combinatorial theory of the Versatile Block we obtain several interlocking assemblies with key differences. The question arises whether there is a possibility to take advantage of the differences to custom-tailor interlocking assemblies for particular applications.

\section*{Contributor Roles}
\noindent\textbf{Tom Goertzen} developed the Versatile Block, wrote code, reviewed the relevant existing literature and wrote a part of this article. \textbf{Domen Macek} planned and performed FEM analyses, validated the numerical results and wrote a part of this article. \textbf{Lukas Schnelle} wrote code, planned and performed FEM analyses and wrote a part of this article.  \textbf{Meike Weiß} planned FEM analyses and wrote part of this article. \textbf{Stefanie Reese}: acquired funding. \textbf{Hagen Holthusen} acquired funding, gave conceptual advice, contributed in the discussion of the results, read the article and gave valuable suggestions for improvement. \textbf{Alice C. Niemeyer} wrote part of this article, acquired funding, gave conceptual advice and supervised.

\section*{Acknowledgements}
The authors gratefully acknowledge funding by the Deutsche Forschungsgemeinschaft (DFG, German Research Foundation) in the framework of the Collaborative Research Centre CRC/TRR 280 “Design Strategies for Material-Minimized Carbon Reinforced Concrete Structures – Principles of a New Approach to Construction” (project ID 417002380).
The authors also thank Max Horn and Katharina Martin for their very valuable comments and advice. 
\newpage

\bibliographystyle{plainnat}
\bibliography{references,Interlocking}

\appendix
\section{Software usage}\label{sec:software}
The interlocking assemblies are generated using the SimplicialSurfaces Package \cite{SimpSurf21} for GAP \cite{GAP4.12}. In this, we generate the different assemblies by first defining and rotating a single Versatile block to all four orientations that can occur in a planar assembly.
Then we create the assemblies given by the wallpaper groups of the three interlockings as described in \ref{subs:ass-vers-block}.\\
Here we divide the assembly into the outermost perimeter of blocks that we use as the frame. In the following the remaining part will be referred to as the core. \\
These assemblies are then exported as individual .stl files, which after a second conversion to .step files can be imported as geometries into abaqus. This is done individually to allow easier applying of the boundary conditions but has no effect on the geometry as a whole. We use a $10 \times 10$ grid of $100$ blocks in each of the periodic interlockings. This is done to have an as similar as possible size of interlocking to compare.

\textit{Note:} The simulations where run with a scaling factor to the coordinates above to fit existing concrete blocks and make it comparable. The scaling was $\frac{0.4}{\sqrt{2}}$ in $x$ and $y$ direction as well as $\frac{1}{5}$ in $z$ direction.

\end{document}